\newif\ifsubmode
\submodefalse

\newif\ifprintfig
\printfigtrue

\ifsubmode
  \documentclass{aastex}
  \usepackage{epsf}
  \received{} 
  \revised{}
  \accepted{}
  \ccc{}
  \cpright{}{}
\else
  \documentclass[11pt,preprint]{aastex}
  \usepackage{epsf}
  \slugcomment{{\it submitted to The Astronomical Journal, March 2000}}
\fi

\shortauthors{Verdoes Kleijn et al.}
\shorttitle{A black hole in IC 1459}

\newcommand{\etal}{{et al.~}}

\newcommand{\lta}{\lesssim}
\newcommand{\gta}{\gtrsim}
\newcommand{\grad}{^{\circ}}
\newcommand{\kms}{\>{\rm km}\,{\rm s}^{-1}}
\newcommand{\ergscm}{\>{\rm erg}\,{\rm s}^{-1}\,{\rm cm}^{-2}}
\newcommand{\pc}{\>{\rm pc}}
\newcommand{\Mpc}{\>{\rm Mpc}}
\newcommand{\Msun}{\>{\rm M_{\odot}}}
\newcommand{\Lsun}{\>{\rm L_{\odot}}}

\newcommand{\Mbh}{M_{\bullet}}
\newcommand{\Rt}{R_{\rm t}}
\newcommand{\Vc}{V_{\rm c}}

\newcommand{\Hbeta}{H$\beta$}
\newcommand{\OIII}{[OIII]}
\newcommand{\OI}{[OI]}
\newcommand{\HalphaNII}{H$\alpha$+[NII]}

\newcommand{\NII}{[NII]}
\newcommand{\SII}{[SII]}

\begin{document}

\title{The black hole in IC 1459\\
from HST observations of the ionized gas disk\altaffilmark{1}}

\author{%
Gijs A.~Verdoes Kleijn,\altaffilmark{2,3}
Roeland P.~van der Marel,\altaffilmark{3}\\
C.~Marcella Carollo,\altaffilmark{4,5}
P.~Tim de Zeeuw\altaffilmark{2}
}

\altaffiltext{1}{Based on observations with the NASA/ESA Hubble Space 
       Telescope obtained at the Space Telescope Science Institute, which is 
       operated by the Association of Universities for Research in Astronomy, 
       Incorporated, under NASA contract NAS5-26555.}

\altaffiltext{2}{Sterrewacht Leiden, Postbus 9513, 2300 RA Leiden, 
The Netherlands.}

\altaffiltext{3}{Space Telescope Science Institute, 3700 San Martin Drive, 
Baltimore, MD 21218.}

\altaffiltext{4}{Columbia University, Department of Astronomy,
538 West 120th Street, New York, NY 10027.}

\altaffiltext{5}{Hubble Fellow.}


\ifsubmode\else
\clearpage\fi


\ifsubmode\else
\baselineskip=14pt
\fi


\begin{abstract}
The peculiar elliptical galaxy IC 1459 ($M_V = -21.19$, $D = 16.5
h^{-1}\Mpc$) has a fast counterrotating stellar core, stellar shells
and ripples, a blue nuclear point source and strong radio core
emission. We present results of a detailed HST study of IC 1459, and
in particular its central gas disk, aimed a constraining the central
mass distribution. We obtained WFPC2 narrow-band imaging centered on
the {\HalphaNII} emission lines to determine the flux distribution of
the gas emission at small radii, and we obtained FOS spectra at six
aperture positions along the major axis to sample the gas
kinematics. We construct dynamical models for the {\HalphaNII} and
{\Hbeta} kinematics that include a supermassive black hole, and in
which the stellar mass distribution is constrained by the observed
surface brightness distribution and ground-based stellar
kinematics. In one set of models we assume that the gas rotates on
circular orbits in an infinitesimally thin disk. Such models
adequately reproduce the observed gas fluxes and kinematics. The
steepness of the observed rotation velocity gradient implies that a
black hole must be present. There are some differences between the
fluxes and kinematics for the various line species that we observe in
the wavelength range 4569 {\AA} to 6819 {\AA}. Species with higher
critical densities generally have a flux distribution that is more
concentrated towards the nucleus, and have observed velocities that are
higher. This can be attributed qualitatively to the presence of the
black hole. There is some evidence that the gas in the central few
arcsec has a certain amount of asymmetric drift, and we therefore
construct alternative models in which the gas resides in collisionless
cloudlets that move isotropically. All models are consistent with a
black hole mass in the range $\Mbh=1$---$4 \times 10^8 \Msun$, and
models without a black hole are always ruled out at high confidence.
The implied ratio of black holes mass to galaxy mass is in the range
$0.4$--$1.5 \times 10^{-3}$, which is not inconsistent with results
obtained for other galaxies. These results for the peculiar galaxy IC
1459 and its black hole add an interesting data point for studies on
the nature of galactic nuclei.
\end{abstract}



\clearpage


\section{Introduction}
\label{s:intro}

Supermassive central black holes (BH) have now been discovered in more
than a dozen nearby galaxies (e.g., Kormendy \& Richstone 1995; Ford
\etal 1998; Ho 1998; Richstone 1998, and van der Marel 1999a for recent
reviews). BHs in quiescent galaxies were mainly found using stellar
kinematics while the BHs in active galaxies were detected through the
kinematics of central gas disks. Other techniques deployed are VLBI
observations of water masers (e.g., Miyoshi \etal 1995) and the
measurement of stellar proper motions in our own Galaxy (Genzel \etal
1997; Ghez \etal 1998). The BH masses measured in active galaxies are
all larger than a few times $10^8 \Msun$, while the BH masses in
quiescent galaxies are often smaller. The number of accurately
measured BHs is expected to increase rapidly in the near future,
especially through the use of STIS on board HST. This will establish
the BH `demography' in nearby galaxies, yielding BH masses as function
of host galaxy properties. In this respect two correlations in
particular have been suggested in recent years. First, a correlation
between BH mass and host galaxy (spheroid) optical luminosity (or
mass) was noted (e.g., Kormendy \& Richstone 1995; Magorrian \etal
1998; van der Marel 1999b). However, this correlation shows
considerable scatter (a factor $\sim 10$ in BH mass at fixed
luminosity). The scatter might be influenced by selection effects
(e.g., it is difficult to detect a low mass BH in a luminous galaxy)
and differences in the dynamical modeling. Second, a correlation
between BH mass and either core or total radio power of the host
galaxy was proposed (Franceschini, Vercellone, \& Fabian
1998). However, the available sample is still small and incomplete.
Establishing the BH masses for a large range of optical and radio
luminosities is crucial to determine the nature of galactic nuclei. An
accurate knowledge of BH demography will put constraints on the
connection between BH and host galaxy formation and evolution and the
frequency and duration of activity in galaxies harboring BHs.

In this paper we measure the BH mass of IC 1459. IC 1459 is an E3
giant elliptical galaxy and member of a loose group of otherwise
spiral galaxies. It is at a distance of $16.5 h^{-1}\Mpc$ with
$M_V=-21.19$ (Faber \etal 1989). Williams \& Schwarzschild (1979)
noted twists in the outer optical stellar isophotes. Stellar spiral
`arms' outside the luminous stellar part of the galaxy were detected
in deep photographs (Malin 1985). Several stellar shells at tens of
kpc from the center were discovered by Forbes \& Reitzel (1995). A
remarkable feature is the counter-rotating stellar core (Franx \&
Illingworth 1988) with a maximum rotation of $\sim 170 \kms$. IC 1459
also has an extended emission gas disk (diameter $\sim 100''$) with
spiral arms (Forbes \etal 1990, Goudfrooij \etal 1990) aligned with
the galaxy major axis. The disk rotates in the same direction as the
outer part of the galaxy (Franx \& Illingworth 1988). The nuclear
region of IC 1459 has line ratios typical of the LINER class (see
e.g., Heckman 1980, Osterbrock 1989 for the definition of LINERS). A
warped dust lane is also present. It is misaligned by $\sim 15\grad$
from the galaxy major axis and some dust patches are observed at a
radius of $2''$ (Carollo \etal 1997). IC 1459 has a blue nuclear
optical source with $V=18.3$ (Carollo \etal 1997; Forbes \etal 1995)
which is unresolved by HST. It also has a variable compact radio core
(Slee \etal 1994). There is no evidence for a radio-jet down to a
scale of $1''$ (Sadler \etal 1989). IC 1459 has a hard X-ray
component, with properties typical of low-luminosity AGNs (Matsumoto
\etal 1997). 

Given the abovementioned properties, IC 1459 might best be described
as a galaxy in between the classes of active and quiescent
galaxies. This makes it an interesting object for extending our
knowledge of BH demography, in particular since there are only few
other galaxies similar to IC 1459 for which an accurate BH mass
determination is available. We therefore made IC 1459, and in
particular its central gas disk, the subject of a detailed study with
the Hubble Space Telescope (HST). We observed the emission gas of IC
1459 with the Second Wide Field and Planetary Camera (WFPC2) through a
narrow-band filter around {\HalphaNII} and took spectra with the Faint
Object Spectrograph (FOS) at six locations in the inner $1''$ of the
disk. In Section~\ref{s:wfpc2} we discuss the WFPC2 observations and
data reduction. In Section~\ref{s:spec} we describe the FOS
observations and data reduction, and we present the inferred gas
kinematics. To interpret the data we construct detailed dynamical
models for the kinematics of the {\Hbeta} and {\HalphaNII} emission
lines in Section~\ref{s:modelH}, which imply the presence of a central
BH with mass in the range $1$---$4 \times 10^8 \Msun$. In
Section~\ref{s:species} we discuss how the kinematics of other
emission line species differ from those for {\Hbeta} and {\HalphaNII},
and what this tells us about the central structure of IC 1459. In
Section~\ref{s:starkin} we present dynamical models for ground-based
stellar kinematical data of IC 1459, for comparison to the results
inferred from the HST data. We summarize and discuss our findings in
Section~\ref{s:discon}.

We adopt $H_0 = 80 \kms \Mpc^{-1}$ throughout this paper. This does
not directly influence the data-model comparison for any of our
models, but does set the length, mass and luminosity scales of the
models in physical units. Specifically, distances, lengths and masses
scale as $H_0^{-1}$, while mass-to-light ratios scale as $H_0$.

\section{Imaging}
\label{s:wfpc2}

\subsection{WFPC2 Setup and Data Reduction}
\label{ss:wfpc2_red}

We observed IC 1459 in the context of HST program GO-6537. We used the
WFPC2 instrument (described in, e.g., Biretta \etal 1996) on September
20, 1996 to obtain images in two narrow-band filters. The observing log
is presented in Table~\ref{t:WFPC2}. The `Linear Ramp Filters' (LRFs)
of the WFPC2 are filters with a central wavelength that varies as a
function of position on the detector. The LRF FR680P15 was used as `on
band' filter, with the galaxy position chosen so as to center the
filter transmission on the {\HalphaNII} emission lines. The narrow-band
filter F631N was chosen as `off-band' filter, and covers primarily
stellar continuum\footnote{The F631N filter covers some of the
redshifted [OI]6300 emission, but the equivalent width of this line is
small enough to have negligible influence on the off-band
subtraction.}. The position of the galaxy on the chip in the off-band
observations was chosen to be the same as in the on-band
observations. In all images the galaxy center was positioned on the PC
chip, yielding a scale of $0.0455''$/pixel.

The images were calibrated with the standard WFPC2 `pipeline', using
the most up to date calibration files. This reduction includes bias
subtraction, dark current subtraction and flat-fielding. A flatfield
was not available for the LRF filter, so we used the flatfield of
F658N, a narrow-band filter with similar central wavelength (6590
\AA). Three back to back exposures were taken through each filter. In
each case, the third exposure was offset by (2,2) PC pixels to
facilitate bad pixel removal. The alignment of the exposures (after
correction for intentional offsets) was measured using both foreground
stars and the galaxy itself, and was found to be adequate.  For each
filter we combined the three available images without additional
shifts, but with removal of cosmic rays, bad pixels and hot pixels.

Construction of a {\HalphaNII} emission image requires subtraction of
the stellar continuum from the on-band image. To this end we first
fitted isophotes to estimate the ratio of the stellar continuum flux
in the on-band and off-band image. This ratio could be fitted as a
slowly varying linear function of radius in regions with no emission
flux. The off-band image was multiplied by this ratio and subtracted
from the on-band image. The resulting {\HalphaNII} emission image was
calibrated to units of $\ergscm$ using calculations with the
STSDAS/SYNPHOT package in IRAF. The resulting flux scale was found to
be in agreement with that inferred from our FOS spectra (see
Section~\ref{s:spec}).

\subsection{The Ionized Gas Disk}
\label{ss:wfpc2_ana}

Figure~\ref{f:images} shows both the F631N stellar continuum image of
the central region of IC1459, as well as the {\HalphaNII} image. The
continuum image shows a weakly obscuring warped dust lane across the
center which is barely visible in the emission image. This dust lane
is evident even more clearly in a $V-I$ image of IC 1459 (Carollo
\etal 1997). The lane makes an angle of $\sim 15 \grad$ with the
stellar major axis.

The {\HalphaNII} emission image shows the presence of a gas disk. The
existence of this disk was already known from ground-based imaging,
which showed that it has a total linear extent of $\sim 100''$
(Goudfrooij \etal 1990; Forbes \etal 1990). The outer parts of the
disk show weak spiral structure and dust patches. Inside the central
$10''$ the disk has a somewhat irregular non-elliptical distribution,
with filaments extending in various directions. Our HST image shows
that the distribution becomes more regular again in the central $r
\lta 0.5''$. Throughout its radial extent, the position angle (PA) 
of the disk coincides with the PA of the stellar distribution. In the
case of the central $0.5''$, we derive from isophotal fits ${\rm PA}
\approx 37\grad$ for the gas disk. This agrees roughly with the PA of
the major axis of the stellar continuum in the same region, for which
the F631N image yields ${\rm PA} \approx 34\grad$. Assuming an
intrinsically circular disk, Forbes \& Reitzel (1995) infer from the
ellipticity $\epsilon = 0.5$ of the gas disk at several arcseconds an
inclination of $60\grad$. We performed a fit to the contour levels of
the extended gas emission published by Goudfrooij \etal (1994), which
also yields an inclination of $60\grad$. By contrast, the gas
distribution in the central $0.5''$ of the HST image is rounder than
that at large radii. The ellipticity increases from $\epsilon = 0.37$
at $r = 1.0''$ to $\epsilon = 0.17$ at $r=0.25''$ (approximately the
smallest radius at which the ellipticity is not appreciably influenced
by the HST point spread function (PSF)). While this could possibly
indicate a change in the inclination angle of the disk, it appears
more likely that the gas disk becomes thicker towards the center. This
latter interpretation receives support from an analysis of the gas
kinematics, as we will discuss below (see
Section~\ref{ss:CTIOcomp}). In the following we therefore assume an
inclination angle of $60\grad$ for IC 1459, as suggested by the
ellipticity of the gas disk at large radii.

\subsection{The Stellar Luminosity Density}
\label{ss:mass_stars}

For the purpose of dynamical modeling we need a model for the stellar
mass density of IC 1459. Carollo \etal (1997) obtained a HST/WFPC2
F814W (i.e., $I$-band) image of IC 1459, and from isophotal fits they
determined the surface brightness profile reproduced in
Figure~\ref{f:sbprof}. Carollo \etal corrected their data
approximately for the effects of dust obscuration through use of the
observed $V-I$ color distribution, so dust is not an important factor
in the following analysis. To fit the observed surface brightness
profile we adopt a parameterization for the three-dimensional stellar
luminosity density $j$. We assume that $j$ is oblate axisymmetric,
that the isoluminosity spheroids have constant flattening $q$ as a
function of radius, and that $j$ can be parameterized as
\begin{equation}
  j(R,z) = j_0 (m/a)^{\alpha} [1+(m/a)^2]^{\beta} , \quad
  m^2 \equiv R^2 + z^2 q^{-2} .
\label{e:lumdendef}
\end{equation}
Here $(R,z)$ are the usual cylindrical coordinates, and $\alpha$,
$\beta$, $a$ and $j_0$ are free parameters. When viewed at inclination
angle $i$, the projected intensity contours are aligned concentric
ellipses with axial ratio $q'$, with $q'^2 \equiv \cos^2 i + q^2
\sin^2 i$. The projected intensity for the luminosity density $j$ 
is evaluated numerically.

In the following we adopt $i=60^{\circ}$, based on the discussion in
Section~\ref{ss:wfpc2_ana}. We take $q'=0.74$ based on the isophotal
shape analysis of Carollo \etal (their figure 1o), which shows an
ellipticity $\epsilon$ of $0.26$ in the inner $10''$ with variations
$< 0.05$. The isophotal PA is almost constant, with a monotonic
increase of $\sim 5^{\circ}$ between $1''$ and $10''$. The Carollo
\etal results show larger variations in $\epsilon$ and PA in the inner
$1''$, but these are probably due to the residual effects of dust
obscuration. Our model with constant $\epsilon$ and PA is therefore
expected to be adequate in the present context.  The projected
intensity of the model was fit to the observed surface brightness
profile between $0.15''$ and $5''$. The best fit model has
$\alpha=-0.62$, $\beta=-0.76$, $a=0.90''$ and $j_0=1.9
\times 10^2 \Lsun \pc^{-3}$. Its predictions are shown by the solid
curve in Figure~\ref{f:sbprof}.

The fit was restricted to the range $R \leq 5''$. As a result, the fit
is somewhat poor at larger radii. This can of course be improved by
extending the fit range, but with the simple parametrization of
equation~(\ref{e:lumdendef}) this would have led to a poorer fit in
the region $R \leq 1''$. Since this is the region of primary interest
in the context of our spectroscopic HST data (described below), we
chose to accept the fit shown in the figure. The central $0.15''$ were
excluded from the fit because IC 1459 has a nuclear point source. This
point source has a blue $V-I$ color. It is most likely of non-thermal
origin (similar to the point source in M87; Kormendy 1992; van der
Marel 1994) and associated with the core radio emission in IC 1459. If
so, the point source does not contribute to the mass density of the
galaxy (which is what we are interested in here), and it is therefore
appropriate to exclude it from consideration.  In
Section~\ref{s:discon} we briefly discuss the implications of the
alternative possibility that the point source is a cluster of young
stars.

\section{Spectroscopy}
\label{s:spec}

\subsection{FOS Setup and Data Reduction}
\label{ss:spec_red}

We used the red side detector of the FOS (described in, e.g., Keyes
\etal 1995) on November 30, 1996 to obtain spectra of IC 1459. The 
COSTAR optics corrected the spherical aberration of the HST primary
mirror. The observations started with a `peak-up' target acquisition
on the galaxy nucleus. The sequence of peak-up stages was similar to
that described in van der Marel, de Zeeuw \& Rix (1997) and van der
Marel \& van den Bosch (1998; hereafter vdMB98). We then obtained six
spectra, three with the {\tt 0.1-PAIR} square aperture (nominal size,
$0.086''$) and three with the {\tt 0.25-PAIR} square aperture (nominal
size, $0.215''$). The G570H grating was used in `quarter-stepping'
mode, yielding spectra with 2064 pixels covering the wavelength range
from 4569 {\AA} to 6819 {\AA}. Periods of Earth occultation were used
to obtain wavelength calibration spectra of the internal arc lamp. At
the end of the observations FOS was used in a special mode to obtain
an image of the central part of IC 1459, to verify the telescope
pointing.

Galaxy spectra were obtained on the nucleus and along the major axis.
A log of the observations is provided in
Table~\ref{t:FOSsetup}. Target acquisition uncertainties and other
possible systematic effects caused the aperture positions on the
galaxy to differ slightly from those commanded to the telescope. We
determined the actual aperture positions from the data themselves,
using the independent constraints provided by the target
acquisition data, the FOS image, and the ratios of the continuum and
emission-line fluxes observed through different apertures. This
analysis was similar to that described in Appendix~A of vdMB98. The
inferred aperture positions are listed in Table~\ref{t:FOSsetup}, and
are accurate to $\sim 0.02''$ in each coordinate. The roll angle of
the telescope during the observations was such that the sides of the
apertures made angles of $39^{\circ}$ and $129^{\circ}$ with respect
to the galaxy major axis. Figure~\ref{f:aperpos} shows a schematic
drawing of the aperture positions. Henceforth we use the labels
`S1'--`S3' for the small aperture observations, and `L1'--`L3' for the
large aperture observations.

Most of the necessary data reduction steps were performed by the HST
calibration pipeline, including flat-fielding and absolute sensitivity
calibration. We did our own wavelength calibration using the arc lamp
spectra obtained in each orbit, following the procedure described in
van der Marel (1997). The relative accuracy (between different
observations) of the resulting wavelength scale is $\sim 0.04${\AA}
($\sim\!2\kms$). Uncertainties in the absolute wavelength scale are
larger, $\sim\!0.4${\AA} ($\sim\!20\kms$), but influence only the
systemic velocity of IC 1459, not the inferred BH mass.

\subsection{Gas Kinematics}
\label{ss:spec_ana}

The spectra show several emission lines, of which the following have a
sufficiently high signal-to-noise ratio ($S/N$) for a kinematical
analysis: {\Hbeta} at 4861 {\AA}; the {\OIII} doublet at 4959, 5007
{\AA}; the {\OI} doublet at 6300, 6364 {\AA}; the {\HalphaNII} complex
at 6548, 6563, 6583 {\AA}; and the [SII] doublet at 6716, 6731 {\AA}.
To quantify the gas kinematics we fitted the spectra under the
assumption that each emission line is a Gaussian. This yields for each
line the total flux, the mean velocity $V$ and the velocity dispersion
$\sigma$. For doublets we fitted both lines simultaneously under the
assumption that the individual lines have the same $V$ and
$\sigma$. The {\HalphaNII} complex is influenced by blending of the
lines, and for this complex we made the additional assumptions that
H$\alpha$ and the [NII] doublet have the same kinematics, and for the
[NII] doublet that the ratio of the fluxes of the individual lines
equals the ratio of their transition probabilities (i.e., 3).

Figure~\ref{f:emlines} shows the observed spectra for each of the five
line complexes listed above, with the Gaussian fits overplotted.  The
figure shows that the observed emission lines are not generally
perfectly fit by Gaussians; they often have a narrower core and
broader wings. It was shown in vdMB98 that this arises naturally in
dynamical models such as those constructed below. In the present paper
we will not revisit the issue of line shapes, but restrict ourselves
to Gaussian fits (both for the data and for our models). The mean and
dispersion of the best-fitting Gaussian are well-defined and
meaningful kinematical quantities, even if the lines themselves are
not Gaussians.

The Gaussian fit parameters for each of the line complexes are listed
in Table~\ref{t:FOSgaskin}. The listed velocities are measured with
respect to the systemic velocity of IC 1459. The systemic velocity was
estimated from the HST data themselves, by including it as a free
parameter in the dynamical models described below (see
Section~\ref{ss:dyn_mod}). This yields $v_{\rm sys} = 1783 \pm 10
\kms$ (but with the possibility of an additional systematic error
due to uncertainties in the FOS absolute wavelength calibration). This
result is a bit higher than values previously reported in the
literature (e.g., $v_{\rm sys} = 1707 \pm 40 \kms$ by Sadler 1984;
$v_{\rm sys} = 1720 \kms$ by Franx \& Illingworth 1988; $v_{\rm sys} =
1748 \pm 42 \kms$ by Da Costa \etal 1991). In fact, systemic
velocities that are up to $150 \kms$ smaller than our value have been
reported as well (e.g., Davies \etal 1987; Drinkwater \etal 1997).

Figure~\ref{f:gaskin} shows the inferred kinematical quantities for
the five line complexes as function of major axis distance. The
observational setup provides only sparse sampling along the major axis
and with apertures of different sizes, but nonetheless, two items are
clear. First, for all apertures and line species there is a steep
positive mean velocity gradient across the nucleus (i.e., between
observations S1 and S2, or L1 and L2). Second, the velocity dispersion
tends to be highest for the smallest aperture closest to the nucleus
(observation S1); this is true for all line species with the exception
of [SII], for which the dispersion peaks for observation S2. The steep
central velocity gradient and centrally peaked velocity dispersion
profile are similar to what has been found for other galaxies with
nuclear gas disks (e.g., Ferrarese, Ford \& Jaffe, 1996; Macchetto
\etal 1997; Bower \etal 1998; vdMB98).

The kinematical properties of the different emission line species show
both significant similarities and differences. For example, the
kinematics of {\Hbeta} and {\HalphaNII} are in excellent quantitative
agreement. By contrast, {\OIII} shows a significantly steeper central
mean velocity gradient, and both {\OIII} and {\OI} have a higher
velocity dispersion for several apertures; the central velocity
dispersion for {\OIII} exceeds that for {\Hbeta} and {\HalphaNII} by
more than a factor two. The kinematics of the {\SII} emission lines
deviates somewhat from that for {\Hbeta} and {\HalphaNII}, but only
for the small apertures. There is no a priori reason to expect
identical flux distributions, and hence identical kinematics for the
different species, because they differ in their atomic structure,
ionization potential, critical density, etc. Differences of similar
magnitude have been detected in the kinematics of other gas disks as
well (e.g., Harms \etal 1994; Ferrarese, Ford \& Jaffe 1996). The
former authors studied the gas disk in M87, and also found that the
{\OIII} line indicates a larger mean velocity gradient and higher
dispersion than {\Hbeta} and {\HalphaNII}. We discuss the differences
in the kinematics of the different line species in
Section~\ref{s:species}, after first having analyzed in detail the
kinematics of {\Hbeta} and {\HalphaNII} in Section~\ref{s:modelH}.

\subsection{Ground-based Spectroscopy}
\label{ss:spec_ground}

In our modeling it proved useful to complement the FOS spectroscopy
with ground-based data that extends to larger radii. We therefore
reanalyzed a major axis long-slit spectrum of IC 1459 obtained at the
CTIO 4m telescope. The data were taken with a $1.5''$-wide slit using
a CCD with $0.73''$ pixels, in seeing conditions with FWHM $\sim
1.9''$. The spectra have a smaller spectral range than the FOS
spectra, but do cover the emission lines of {\Hbeta} and
{\OIII}. Fluxes and kinematics for these lines were derived using
single Gaussian fits, as for the FOS spectra. The inferred gas
kinematics are listed in Table \ref{t:CTIO}. The stellar kinematics
implied by the {\it absorption} lines in the same spectrum (presented
previously by van der Marel \& Franx 1993) are used in
Section~\ref{s:starkin} for the construction of stellar dynamical
models.

\section{Modeling and Interpretation of the {\HalphaNII} and 
{\Hbeta} Kinematics}
\label{s:modelH}

The FOS spectra of the {\HalphaNII} and {\Hbeta} lines yield similar
relative fluxes (cf.~Figure~\ref{f:fluxfit} below) and similar mean
velocities and dispersions (cf.~Figure~\ref{f:gaskin}). We therefore
assume that these emission lines have the same intrinsic flux
distributions and kinematics. We start in
Sections~\ref{ss:flux_gas}--\ref{ss:CTIOcomp} with the construction of
models for the {\Hbeta} and {\HalphaNII} gas kinematics in which the
gas disk is assumed to be an infinitesimally thin structure in the
equatorial plane of the galaxy. However, as discussed in
Section~\ref{ss:wfpc2_ana}, this assumption may not be entirely
appropriate at small radii, where the projected isophotes of the gas
disk become rounder. In Section~\ref{ss:asymdrift} we therefore
discuss models in which the gas distribution is extended vertically.
 
\subsection{Flux Distribution}
\label{ss:flux_gas}

To model the {\Hbeta} and {\HalphaNII} gas kinematics we need a
description of the intrinsic (i.e., the deconvolved and de-inclined)
flux profile for these emission lines. We model the (face-on)
intrinsic flux distribution as a triple exponential,
\begin{equation}
  F(R) = F_1 \exp(-R/R_1) + F_2 \exp(-R/R_2) + F_3 \exp(-R/R_3) ,
\label{e:fluxparam}
\end{equation}
and assume that the disk is infinitesimally thin and viewed at an
inclination $i=60\grad$, (cf.~Section~\ref{ss:wfpc2_ana}). The total
flux contributed by each of the three exponential components is $I_i =
2 \pi R_i^2 F_i$ ($i=1,2,3$), and the overall total flux is $I_{\rm
tot} = I_1 + I_2 + I_3$.

The best-fitting parameters of the model flux distribution were
determined by comparison to the available data. Flux data are
available for {\HalphaNII} from both the WFPC2 imaging and FOS
spectra. For {\Hbeta} they are available from the FOS and the CTIO
spectra. For the spectra we determined the fluxes in the relevant
lines (and their formal errors) using single Gaussian
fits.\footnote{It was verified that the fluxes extracted using single
Gaussian fits are not significantly different from those obtained by
simply adding the pixel data in the relevant wavelength range.} For
the WFPC2 image data we included for simplicity not the full
two-dimensional brightness distribution in the fit, but only image
cuts along the major and minor axes. Image fluxes outside $\sim 1''$
are dominated by read-noise, and were excluded. The errors for each
image data-point were computed taking into account the Poisson-noise
and the detector read-noise. The combined flux data from all sources
are shown in Figure~\ref{f:fluxfit}.

We performed an iterative fit of the triple exponential to all the
available flux data, taking into account the necessary convolutions
with the appropriate PSF, pixel size and aperture size for each
setup. The HST and CTIO fluxes constrain the flux distribution
predominantly for $r \lta 1''$ and $r \gta 1''$, respectively, due to
their relatively narrow and broad PSF. The WFPC2 data have a pixel
area that is $\sim 3$ and $\sim 20$ times smaller than the respective
FOS apertures, and therefore provide the strongest constraints on the
flux distribution close to the center.  The solid line in
Figure~\ref{f:fluxfit} shows the predictions of the model that best
fits all available data (which we will refer to as `the standard flux
model'). This model has parameters $R_1 = 0.026''$, $R_2 = 0.20''$,
$R_3 = 1.65''$, $I_1 / I_{\rm tot} = 0.260$, $I_2 / I_{\rm tot} =
0.353$ and $I_3 / I_{\rm tot} = 0.387$. The absolute calibration gives
$I_{\rm tot} = 3.0 \times 10^{-13} \ergscm$ for {\HalphaNII} and
$I_{\rm tot} = 1.3 \times 10^{-14} \ergscm$ for {\Hbeta}. The total
{\HalphaNII} flux inferred from our model agrees to within 25\% with
that inferred from a previous ground-based observation of IC 1459
(Macchetto \etal 1996). Figure~\ref{f:fluxprof} shows the intrinsic
flux distribution as function of radius for the standard flux
model. Approximately one quarter of the total flux is contained in a
component that is essentially unresolved at the spatial resolution of
HST.

The standard flux model provides an adequate fit to the observed
fluxes, but the fit is not perfect. The model predicts too little flux
in the central WFPC2 pixel, while at the same time predicting too much
flux in the small FOS aperture closest to the galaxy center.  So the
different data sets are not fully mutually consistent under the
assumptions of our model. This is presumably a result of uncertainties
in the PSFs and aperture sizes for the different observations. To
explore the influence of this on the inferred flux distribution we
performed fits to two subsets of the flux data. The first subset
consists only of the FOS and CTIO data, while the second subset
consists only of the WFPC2 and CTIO data. The flux distribution models
that best fit these subsets of the data are also shown in
Figure~\ref{f:fluxprof}. The results show that the standard flux model
represents a compromise between the FOS and the WFPC2 data. At small
radii the FOS data by themselves would imply a broader profile, while
the WFPC2 data by themselves would imply a narrower profile. At larger
radii the situation reverses. As will be discussed in
Section~\ref{ss:bestfit}, the uncertainties in the intrinsic
flux distribution have only a very small effect on the inferred BH
mass.

\subsection{Dynamical Models}
\label{ss:dyn_mod}

Our thin-disk models for the gas kinematics are similar to those
employed in vdMB98. The galaxy model is axisymmetric, with the stellar
luminosity density $j(R,z)$ chosen as in Section~\ref{ss:mass_stars}
to fit the available surface photometry. The stellar mass density
$\rho(R,z)$ follows from the luminosity density upon the assumption of
a constant mass-to-light ratio $\Upsilon$. The mass-to-light ratio can
be reasonably accurately determined from the ground-based stellar
kinematics for IC 1459. This yields $\Upsilon = 4.5$ in solar $I$-band
units, cf.~Section~\ref{s:starkin} below. We keep the mass-to-light
ratio fixed to this value in our modeling of the gas kinematics. We
assume that the gas is in circular motion in an infinitesimally thin
disk in the equatorial plane of the galaxy, and has the circularly
symmetric flux distribution $F(R)$ given in
Section~\ref{ss:flux_gas}. We take the inclination of the galaxy and
the gas disk to be $i=60\grad$, as discussed in
Section~\ref{ss:wfpc2_ana}.  The circular velocity $\Vc(R)$ is
calculated from the combined gravitational potential of the stars and
a central BH of mass $\Mbh$.  The line-of-sight velocity profile (VP)
of the gas at position $(x,y)$ on the sky is a Gaussian with mean
$\Vc(R) \sin i$ and dispersion $\sigma_{\rm gas}(R)$, where $R =
\sqrt{x^2 + (y/\cos i)^2}$ is the radius in the disk. The velocity
dispersion of the gas is assumed to be isotropic, with contributions
from thermal and non-thermal motions: $\sigma_{\rm gas}^2 =
\sigma_{\rm th}^2 + \sigma_{\rm turb}^2$. We refer to the non-thermal 
contribution as `turbulent', although we make no attempt to describe
the underlying physical processes that cause this dispersion. It
suffices here to parameterize $\sigma_{\rm turb}$ through:
\begin{equation}
  \sigma_{\rm turb}(R) = \sigma_0 + [\sigma_1 \exp(-R/\Rt)] .
\label{eq:turbdef}
\end{equation}
The parameter $\sigma_0$ was kept fixed to $120 \kms$, as suggested by
the CTIO data for H$\beta$ with $|r| \gta 3.5''$ (see
Figure~\ref{f:modelctio} below). The predicted VP for any given
observation is obtained through flux weighted convolution of the
intrinsic VPs with the PSF of the observation and the size of the
aperture. The convolutions are described by the semi-analytical
kernels given in Appendix~A of van der Marel \etal (1997), and were
performed numerically using Gauss-Legendre integration. A Gaussian is
fit to each predicted VP for comparison to the observed $V$ and
$\sigma$.

The model was fit to the FOS gas kinematics for {\HalphaNII} and
{\Hbeta}. Rotation velocity and velocity dispersion measurements were
both included, yielding a total of 24 data points. Three free
parameters are available to optimize the fit: $\Mbh$, and the
parameters $\sigma_1$ and $\Rt$ that describe the radial
dependence of the turbulent dispersion. The temperature of the gas is
not an important parameter: the thermal dispersion for $T \approx 10^4
{\rm K}$ is $\sigma_{\rm th} \approx 10\kms$, and is negligible with
respect to $\sigma_{\rm turb}$ for all plausible models. We define a
$\chi^2$ quantity that measures the quality of the fit to the
kinematical data, and the best-fitting model was found by minimizing
$\chi^2$ using a `downhill simplex' minimization routine (Press \etal
1992).

\subsection{Data-model comparison for the FOS data}
\label{ss:bestfit}

The curves in Figure~\ref{f:modelfos} show the predictions of the
model that provides the overall best fit to the {\HalphaNII} and
{\Hbeta} kinematics, using the standard flux model of
Section~\ref{ss:flux_gas}. Its parameters are: $\Mbh = 1.0
\times 10^8 \Msun$, $\sigma_1 = 563 \kms$ and $\Rt=0.1''$. 
This model (which we will refer to as `the standard kinematical
model') adequately reproduces the important features of the HST
kinematics, including the central rotation gradient and the nuclear
velocity dispersion.

To determine the range of BH masses that provides an acceptable fit to
the data we compared the predictions of models with different fixed
values of $\Mbh$, while at each $\Mbh$ varying the remaining
parameters to optimize the fit. The radial dependence of the intrinsic
velocity dispersion of the gas is essentially a free function in our
models, so the observed velocity dispersion measurements can be fit
equally well for all plausible values of $\Mbh$. Thus only the
predictions for the HST rotation velocity measurements depend
substantially on the adopted $\Mbh$. To illustrate the dependence on
$\Mbh$, Figure~\ref{f:rotfits} compares the predictions for the HST
rotation measurements for three different models. The solid curves
show the predictions of the standard kinematical model defined
above. The dotted and dashed curves are the predictions of models in
which $\Mbh$ was fixed a priori to $0$ and $7.0 \times 10^8 \Msun$,
respectively. The model without a BH predicts a rotation curve slope
which is too shallow and the model with $\Mbh = 7.0 \times 10^8
\Msun$ predicts a rotation curve slope which is too steep. Both 
these BH masses are ruled out by the data at more than the $99\%$
confidence level (see discussion below).

To assess the quality of the fit to the HST rotation velocity
measurements we define a new $\chi^2$ quantity, $\chi^2_V$, that
measures the fit to these data only. At each $\Mbh$, the parameters
$\sigma_1$ and $\Rt$ are fixed almost entirely by the velocity
dispersion measurements. These parameters can therefore not be varied
independently to improve the fit to the HST rotation velocity
measurements. As a result, $\chi^2_V$ is expected to follow
approximately a $\chi^2$ probability distribution with $N_{\rm df} =
12-1 = 11$ degrees of freedom (there are 12 HST measurements, and
there is one free parameter, $\Mbh$). The expectation value for this
distribution is $\langle \chi^2_V \rangle = 11$. However, for the
standard kinematical model we find $\chi^2_V = 59$. To determine
the cause of this statistically poor fit we inspected the goodness of
fit as function of BH mass for each line species separately.

Figure~\ref{f:mbhrange} shows $\chi^2_V$ as function of $\Mbh$ for
both {\HalphaNII} and {\Hbeta}. The kinematics of {\HalphaNII} are
formally poorly fitted, despite the apparently good qualitative
agreement in Figure~\ref{f:modelfos}. In particular, the observed
{\HalphaNII} velocity gradient between the FOS-0.1 apertures S1 and S2
is steeper than predicted by the best-fit model with $\Mbh = 1.0
\times 10^8 \Msun$, which suggests that the BH mass may actually be twice 
as high (since $\Mbh \propto \Delta V^2$ in our models). The poor
formal fit may not be too surprising, given that our modeling of the
gas as a flat circular disk in bulk circular rotation with an
additional turbulent component is almost certainly an
oversimplification of what in reality must be a complicated
hydrodynamical system. The fits to the kinematics of {\Hbeta} are
statistically acceptable, but this may be in part because the formal
errors on the {\Hbeta} kinematics are twice as large as for
{\HalphaNII}. This would cause any shortcomings in the models to be
less apparent for this emission line. Nonetheless, an important result
in Figure~\ref{f:mbhrange} is that the BH masses implied by the
{\HalphaNII} and {\Hbeta} kinematics are virtually identical. Formal
errors on the BH mass can be inferred using the $\Delta\chi^2$
statistic, as illustrated in Figure~\ref{f:mbhrange}. For H$\beta$
this yields $\Mbh = (1.0 \pm 0.5) \times 10^8 \Msun$ at $68.3$\%
confidence (i.e., 1-$\sigma$), and $0.2 \times 10^8 \Msun \leq \Mbh
\leq 2.5 \times 10^8 \Msun$ at 99\% confidence. The formal $\Delta\chi^2$ 
confidence intervals inferred from the {\HalphaNII} lines are smaller,
but this is not necessarily meaningful since the $\chi^2$ itself is
not acceptable for these lines.

In Section~\ref{ss:flux_gas} we showed that there is some uncertainty
in the flux distributions of {\HalphaNII} and {\Hbeta}. The mean
velocities and velocity dispersions predicted by the dynamical model
are flux-weighted quantities, and therefore depend on the adopted flux
distribution. To assess the influence on the inferred BH mass we
repeated the analysis using the two non-standard flux distributions
shown in Figure~\ref{f:fluxprof}. With these distributions we found
fits to the kinematical data of similar quality as for the standard
kinematical model. The inferred values of $\Mbh$ agree with those for
the standard kinematical model to within 10\%. This shows that the
uncertainties in the flux distribution have negligible impact on the
inferred BH mass.

\subsection{Data-model comparison for the CTIO data}
\label{ss:CTIOcomp}

The model parameters in Section~\ref{ss:bestfit} were chosen to best
fit the FOS data. Here we investigate what this model predicts for the
setup of the CTIO data. Figure~\ref{f:modelctio} shows the resulting
data-model comparison (without any further changes to the model
parameters). 

At radii $|r| \gta 4''$ the standard kinematical model fits the data
acceptably well. The agreement in the velocity dispersion is trivial
since it is the direct result of our choice of the model parameter
$\sigma_0$. However, the agreement for the rotation velocities is
quite important. It shows that outside the very center of the galaxy,
the observations are consistent with the assumed scenario of gas
rotating at the circular velocity in an infinitesimally thin
disk. Moreover, it suggests that the value of mass-to-light ratio
$\Upsilon$ used in the models (derived from ground-based stellar
kinematics) is accurate.

By contrast, the fit to the CTIO data is less good at radii $|r| \lta
4''$. In particular, the predicted rotation curve is too steep, and
the central peak in the predicted velocity dispersion is too
high. These discrepancies cannot be attributed to possible errors in
the assumed value for the seeing FWHM for the CTIO observations. The
latter was calibrated from the spectra themselves. Due to their
superior spatial resolution, the WFPC2 data set the inner intrinsic
flux profile. We could therefore determine the CTIO PSF by optimizing
the agreement between the predicted and observed central three CTIO
fluxes using this intrinsic inner flux profile. The resulting FWHM
determination ($1.9''$) is quite accurate, and is inconsistent with
the large FWHM values needed to make the standard kinematical model
fit the CTIO data. The discrepancies must therefore be due to an
inaccuracy or oversimplification in the modeling. 

The models discussed so far assume that the observed velocity
dispersion is due to local turbulence in gas that has bulk motion
along circular orbits. However, an alternative interpretation could be
that the gas resides in individual clouds, and that the observed
dispersion of the gas is due to a spread in the velocities of
individual clouds. In this case the gas would behave as a
collisionless fluid obeying the Boltzmann equation. An important
consequence would be that the velocity dispersion $\sigma$ of the gas
clouds would provide some of the pressure responsible for hydrostatic
support, so that the mean rotation velocity $\overline{v_{\phi}}$
would be less than the circular velocity $v_{\rm c}$ by a certain
amount $\Delta v \equiv v_{\rm c} - \overline{v_{\phi}}$. This effect
is know as asymmetric drift (for historical reasons having to do with
the stellar dynamics of the Milky Way; see e.g., Binney \& Merrifield
1998).

The presence of a certain amount of asymmetric drift would
simultaneously explain several observations. First, if close to the
center the gas receives a certain amount of pressure support from bulk
velocity dispersion, and if this dispersion would be near-isotropic,
it would induce a thickening of the disk. This would cause the
isophotes of the gas close to the center to be rounder than those at
larger radii, which is exactly what is observed
(cf.~Sections~\ref{ss:wfpc2_ana} and \ref{s:discon}). Second,
asymmetric drift would cause the gas to have a mean velocity less than
the circular velocity, which would explain why the models of
Section~\ref{ss:bestfit} overpredict the observed rotation velocities
(cf.~Figure~\ref{f:modelctio}). Third, the central peak in the
velocity dispersion seen in the ground-based data is due in part to
rotational broadening (spatial convolution of the steep rotation
gradient near the center). So if the mean velocity of the gas in the
models were lowered due to asymmetric drift, then the predicted
central velocity dispersion would go down as well. This would tend to
improve the agreement between the predicted and the observed velocity
dispersions in Figure~\ref{f:modelctio}.

\subsection{The influence of asymmetric drift}
\label{ss:asymdrift}

If the gas in IC 1459 indeed has a non-zero asymmetric drift at small
radii, then the models of Section~\ref{ss:dyn_mod} would have {\it
under\/}-estimated the enclosed mass within any given radius (and
models without a BH would be ruled out at even higher confidence than
already indicated by Figure~\ref{f:mbhrange}). In the following we
address how much of an effect this would have on the inferred BH
mass. In the limit $\sigma / \overline{v_{\phi}} \ll 1$ one has that
$\Delta v / v_{\rm c} = {\cal O} ([\sigma/v_{\rm c}]^2)$ (e.g., Binney
\& Tremaine 1987), and any asymmetric drift correction to $\Mbh$ would
be fairly small. However, at the resolution of HST we find that
$\sigma / \overline{v_{\phi}} \gta 1$, so the approximate formulae
that exist for the limit $\sigma / \overline{v_{\phi}} \ll 1$ cannot
be used. For a proper analysis the gas kinematics would have to be
modeled as a `hot' system of point masses, using any of the techniques
that have been developed in the context of stellar dynamical modeling
of elliptical galaxies (e.g., Merritt 1999).

While fully general collisionless modeling of the gas in IC 1459 is
beyond the scope of the present paper, it is important to establish
whether such an analysis would yield a very different BH mass. To
address this issue we constructed spherical isotropic models for the
gas kinematics using the Jeans equations, as in van der Marel
(1994). The three-dimensional density of the gas was chosen so as to
reproduce the major axis profile given by equation~(\ref{e:fluxparam})
after projection. As before, the gravitational potential of the system
was characterized by a variable $\Mbh$ and a fixed $\Upsilon = 4.5$.
Any turbulent velocity dispersion component of the gas was assumed to
be zero. We then calculated the RMS projected line-of-sight velocity
$v_{\rm RMS}$ predicted for the smallest FOS aperture positioned on
the galaxy center. The {\HalphaNII} and {\Hbeta} observations yield
$v_{\rm RMS} \equiv [V^2 + \sigma^2]^{1/2} \approx 600 \kms$
(cf.~Table~\ref{t:FOSgaskin}). We found that the spherical isotropic
Jeans models require $\Mbh = 4.0 \times 10^8 \Msun$ to reproduce this
value. Larger BH masses would be ruled out because they predict more
RMS motion than observed. In more general collisionless models the
required BH mass will depend on the details of the model, but not very
strongly. Velocity anisotropy and axisymmetry influence the projected
dispersion of a population of test particles in a Kepler potential
only at the level of factors of order unity (de Bruijne, van der Marel
\& de Zeeuw 1996).

So as expected, if the velocity dispersion of the gas is interpreted
as gravitational motion of individual clouds, then the BH mass must be
larger than inferred in Section~\ref{ss:bestfit}. However, the
increase would only be a factor of $\sim 3$---4. Models without a BH
would remain firmly ruled out. We emphasize that both types of model
that we have studied are fairly extreme. In one case we assume that
the gas resides in an infinitesimally thin disk, and has a large
turbulent (or otherwise non-thermal) velocity dispersion and no bulk
velocity dispersion or asymmetric drift. In the other case we assume
the opposite, that the gas is in a spherical distribution, and has no
turbulent velocity dispersion but instead a large bulk velocity
dispersion. The truth is likely to be found somewhere between these
extremes, and we therefore conclude that IC 1459 has a BH with mass in
the range $\Mbh = 1$---$4 \times 10^8 \Msun$.

\section{Modeling and Interpretation of the [OIII], [OI] and [SII] data}
\label{s:species}

Spectroscopic information is not only available for {\HalphaNII} and
{\Hbeta}, but also for three other line species: {\OI}, {\OIII} and
{\SII}. Interestingly, the flux distributions and kinematics of these
lines show some notable differences from those of {\Hbeta} and
{\HalphaNII}.

We have no narrow-band imaging for {\OI}, {\OIII} and {\SII}, so
information on the flux distributions of these lines is available only
from the six apertures for which we obtained FOS spectra.
Figure~\ref{f:speciesflux} shows the relative surface brightness
${\cal F}$ for the different apertures and line species, using the
definition
\begin{equation}
  {\cal F} {\rm (aperture,species)} \equiv 
      [ F {\rm (aperture,species)} / A {\rm (aperture)} ] \> \Big / \>
      [ F {\rm (L2,species)} / A {\rm (L2)} ] ,
\label{eq:relfluxdef}
\end{equation}
where $F$ is the flux observed through an aperture for a given
species, and $A$ is the area of the aperture; the aperture L2 is the
observation with the large {\tt 0.25-PAIR} aperture on the nucleus
(cf.~Figure~\ref{f:aperpos}). The relative surface brightnesses for
the different species as seen through the large aperture are all quite
similar. By contrast, those for the small {\tt 0.1-PAIR} aperture
differ considerably. The profiles for {\OI} and especially {\OIII} are
more centrally peaked than that for {\HalphaNII}, while the profile
for {\SII} is less centrally peaked than that for {\HalphaNII}. To
quantify this, we have defined a measure of the peakedness of the
surface brightness profile as
\begin{equation}
  {\cal P} {\rm (species)} \equiv
      2 {\cal F} {\rm (S1,species)} \> \Big / \>
      [ {\cal F} {\rm (S2,species)} + {\cal F} {\rm (S3,species)} ] .
\label{eq:peakedness}
\end{equation}
The values of this quantity for the different species are $2.1$,
$3.5$, $3.0$, $1.9$ and $0.87$ for {\Hbeta}, {\OIII}, {\OI},
{\HalphaNII} and {\SII}, respectively.

The kinematics for the different species also show differences, as
already pointed out in Section~\ref{ss:spec_ana}.
Figure~\ref{f:gaskin} shows that the most pronounced differences are
seen in the value of the velocity dispersion as observed through the
smallest aperture on the nucleus (observation S1). This quantity
varies from $211 \pm 25 \kms$ for {\SII} to $1014 \pm 47 \kms$ for
{\OIII}, cf.~Table~\ref{t:FOSgaskin}. Such differences in the line
width for different species have previously been found in the central
regions of other LINER galaxies and Seyferts, and have been
extensively modeled (e.g., Whittle 1985; Simpson \& Ward 1996; Simpson
\etal 1996; and references therein). One result from these studies has
been that there is generally a correlation between velocity dispersion
and critical density of the lines. We find a similar result for the
nucleus of IC 1459.  Figure~\ref{f:critical}a shows the velocity
dispersion for the different lines observed through aperture S1,
versus the critical density (the Balmer lines are not plotted because
for them interpretation of the critical density is complicated by the
effects of radiative transfer for permitted lines; see e.g.~Filippenko
\& Halpern 1984). There is indeed a rough correlation. The fact that
{\OIII} has a larger dispersion than {\OI} is somewhat surprising in
view of this, but this has also been found for other galaxies (Whittle
1985). It has been hypothesized that at a basic level the approximate
correlation between velocity dispersion and critical density can be
understood as the result of differences in the spatial distribution of
the line flux for different species (Osterbrock 1989, p.~366). Lines
with a high critical density tend to be more strongly concentrated
towards the ionizing source in the galaxy nucleus than species with a
low critical density. So for a line with a high critical density, the
observed flux will on average come from smaller radii. Either the
presence of a central BH or increased turbulence (cf.\
equation~\ref{eq:turbdef}) would naturally cause the gas at smaller
radii to move faster, which qualitatively explains the correlation. To
make detailed quantitative predictions one would need to model the
complete ionization structure (ionizing flux, electron density,
temperature, etc.~as function of radius) and kinematics of the gas,
which can generally be done only with simplifying
assumptions. However, even without a detailed model we can test the
basic interpretation in an observational sense.
Figure~\ref{f:speciesflux} shows that we are observing different flux
distributions for the different lines, which implies that we are
actually resolving the region from which the emission arises. Thus we
can test directly whether the lines for which the velocity dispersion
is high have a flux that is strongly concentrated towards the
nucleus. Figure~\ref{f:critical}b confirms this. It shows the velocity
dispersion for the different lines observed through aperture S1 versus
the flux-peakedness parameter ${\cal P}$. The strong correlation
provides direct support for the proposed interpretation.
 
In our dynamical models, the gas motions are a function of position in
the disk only; they do not depend on the physical properties of the
line species. Hence, differences in the observed kinematics of the
lines must be due entirely to differences in their flux distributions.
For accurate modeling it is therefore essential that the intrinsic
flux distributions are well-constrained by the observations. This is
true for {\HalphaNII}, for which a narrow-band image is available at
the WFPC/PC resolution of $0.0455''$/pixel. However, this is not true
for {\OI}, {\OIII} and {\SII}, for which only the six FOS spectral
flux measurements are available, with resolutions no better than
$0.086''$/aperture. This is insufficient for accurate modeling. Hence,
we cannot test in detail whether the different line species all
independently indicate the same BH mass. However, a very simple
argument can be used to set an upper limit on how much the BH masses
implied by the {\OI}, {\OIII} and {\SII} data could differ from that
inferred in Section~\ref{ss:bestfit} from the {\HalphaNII} and
{\Hbeta} data. None of the line species have either a central rotation
velocity gradient $\Delta V \equiv V({\rm S2}) - V({\rm S1})$ or a
central velocity dispersion $\sigma({\rm S1})$ that exceeds the value
for {\HalphaNII} by more than a factor of $\sim 2$
(cf.~Table~\ref{t:FOSgaskin}). In the models of
Section~\ref{ss:dyn_mod} one has approximately $\Mbh \propto
\Delta V^2$, while in the models of Section~\ref{ss:asymdrift} one has
approximately $\Mbh \propto \sigma({\rm S1})^2$. So if we would assume
(incorrectly) that all species have the same flux distribution, then
we would infer BH masses that exceed that inferred from the
{\HalphaNII} and {\Hbeta} data by at most a factor $\sim 4$. However,
this is a very conservative upper limit since the flux distributions
for e.g. {\OI} and {\OIII} {\it are} actually more peaked than for
{\HalphaNII} and {\Hbeta}, which would tend to reduce the inferred BH
mass.  So the data provide no compelling reason to believe that the
{\OI}, {\OIII} and {\SII} observations imply a very different BH mass
than inferred from {\HalphaNII} and {\Hbeta}, but we cannot test this
in detail.

For {\SII}, both the mean velocity and the velocity dispersion
profiles are quite irregular. The {\SII} lines have a relatively low
equivalent width and form a blended doublet, so this could be due to
systematic problems in the extraction of the kinematics from the data.
On the other hand, irregularities in the dispersion profiles are also
seen for {\OIII} and {\OI} as observed through the small apertures.
Although such irregularities are not seen for {\Hbeta} and
{\HalphaNII}, this may indicate that our modeling of the gas flux
distribution (eq.~[\ref{e:fluxparam}]) and the turbulent velocity
dispersion (eq.~[\ref{eq:turbdef}]) as smooth functions is somewhat
oversimplified. We compared the observed line ratios for IC 1459 with
modeled values for shocks (Allen \etal 1998; Dopita \etal 1997) and
found them to be consistent. Shocks could naturally explain the
turbulence that we invoke in our models of IC 1459. At the same time,
it would suggest that the gas properties could easily possess more
small-scale structure than the smooth functions that we have adopted.

\section{Ground-based Stellar Kinematics}
\label{s:starkin}

Ground-based stellar kinematical data are available from the same
major axis CTIO spectrum for which the emission lines were discussed
in Section~\ref{ss:spec_ground}. The stellar rotation velocities $V$
and velocity dispersions $\sigma$ inferred from this spectrum were
presented previously in van der Marel \& Franx (1993). To interpret
these data we constructed a set of axisymmetric stellar-dynamical
two-integral models for IC 1459 in which the phase-space distribution
function $f(E,L_z)$ depends only on the two classical integrals of
motion. As before, the mass density was taken to be $\rho = \Upsilon
j$, with the luminosity density $j(R,z)$ given by
equation~(\ref{e:lumdendef}). The gravitational potential is the sum
of the stellar potential and the Kepler potential of a possible
central BH. Predictions for the root-mean-square (RMS) stellar
line-of-sight velocities $v_{\rm RMS} \equiv [V^2 + \sigma^2]^{1/2}$
were calculated by solving the Jeans equations for hydrostatic
equilibrium, projecting the results onto the sky, and convolving them
with the observational setup. This modeling procedure is equivalent to
that applied to large samples by, e.g., van der Marel (1991) and
Magorrian \etal (1998). The calculations presented here were done with
the software developed by van der Marel \etal (1994).

Figure~\ref{f:starkin} shows the data for $v_{\rm RMS}$ as function of
major axis distance. The value of $\Upsilon$ in the models was chosen
so as to best fit the data outside the central region, yielding
$\Upsilon \approx 4.5$. This leaves $\Mbh$ as the only remaining free
parameter. The curves in Figure~\ref{f:starkin} show the model
predictions for various values of $\Mbh$. A model without a BH
predicts a central dip in $v_{\rm RMS}$, which is not observed. The
models therefore clearly require a BH. None of the models fits
particularly well, but models with $\Mbh \approx 3$--$4 \times 10^9
\Msun$ provide the best fit. This exceeds the BH mass inferred from
the HST gas kinematics by a factor of 10 or more. This suggests that
the assumptions underlying the stellar kinematical analysis may not be
correct. The velocity dispersion anisotropy of a stellar system can
have any arbitrary value, and the sense of the anisotropy can have a
large effect on inferences about the nuclear mass
distribution. Two-integral $f(E,L_z)$ models can be viewed as
axisymmetric generalizations of spherical isotropic models. Such
models will overestimate the BH mass if galaxies are actually radially
anisotropic (Binney \& Mamon 1982). Van der Marel (1999a) showed that
$f(E,L_z)$ models can easily overestimate the BH mass by a factor of
10 when applied to ground-based data of similar quality as that
available here, even for galaxies that are only mildly radially
anisotropic. Support that this may be happening comes from various
directions. First, several detailed studies of bright galaxies similar
to IC 1459 have concluded that these galaxies are radially anisotropic
(e.g., Rix \etal 1997; Gerhard \etal 1998; Matthias \& Gerhard 1999;
Saglia \etal 1999; Cretton, Rix \& de Zeeuw 2000). Second, detailed
three-integral distribution function modeling of stellar kinematical
HST data for several galaxies (Gebhardt \etal 2000) does indeed yield
BH masses that are many times smaller than inferred by Magorrian
\etal (1998) from $f(E,L_z)$ models for ground-based data for the same 
galaxies. And third, models of adiabatic BH growth for HST photometry
also suggest that $f(E,L_z)$ models for ground-based data yield BH
masses that are too large (van der Marel 1999b). So in summary, the
fact that the BH mass inferred from $f(E,L_z)$ models for ground-based
IC 1459 data does not agree with that inferred from the HST gas
kinematics provides little reason to be worried. It simply shows that
one cannot generally place very meaningful constraints on BH masses
from ground-based stellar kinematics of $\sim 2''$ spatial resolution.

While the BH mass inferred from ground-based stellar kinematical data
is generally very sensitive to assumptions about the structure and
dynamics of the galaxy, this is not true for the inferred
mass-to-light ratio. Models with different inclination (van der Marel
1991) or anisotropy (van der Marel 1994; 1999a) yield the same value
of $\Upsilon$ to within $\sim 20$\%. Additional support for the
accuracy of the inferred $\Upsilon = 4.5$ comes from the fact that
this value yields a good fit to the observed rotation velocities of
the gas outside the central few arcsec (cf.~Figure~\ref{f:modelctio}).

\section{Discussion and conclusions}
\label{s:discon}

We have presented the results from a detailed HST study of the central
structure of IC 1459. The kinematics of the gas disk in IC 1459 was
probed with FOS observations through six apertures along the major
axis. In our modeling of the observed kinematics we took into account
the stellar mass density in the central region by fitting WFPC2
broad-band imaging, and we determined the flux distribution of the
emission-gas from WFPC2 narrow-band imaging. From the models we have
determined that IC1459 harbors a black hole with a mass in the range
$1 \times 10^8 \Msun$ -- $4 \times 10^{8} \Msun$, with the exact value
depending somewhat on whether we model the gas as rotating on circular
orbits, or as an ensemble of collisionless cloudlets. While the
dynamical models that we have constructed provide good fits to the
observations, the true structure of IC 1459 could of course be more
complex than our models. Below we discuss several aspects of this.
  
Ground-based observations (Goudfrooij \etal 1994; Forbes \& Reitzel
1995) indicate an ellipticity $\epsilon = 0.5$ for the gas disk at
radii larger than a few arcseconds, implying an inclination angle
$i=60\grad$. By contrast, from our HST emission-line image we found a
monotonic increase in ellipticity from $\epsilon = 0.17$ to $\epsilon
= 0.37$ between $r=0.25''$ to $1.0''$. In our modeling we have assumed
that these rounder inner isophotes are due to a thickening of the gas
disk caused by asymmetric drift, and we estimated the effect of this
on the inferred $\Mbh$ (see Sections~\ref{ss:CTIOcomp} and
\ref{ss:asymdrift}). However, an alternative interpretation would be
to assume that the disk is warped. The presence of a dust lane
slightly misaligned with the gas disk, a counter-rotating stellar
core, and stellar shells and ripples in the outer galaxy make it
plausible that the central gas and dust were accreted from outside,
displaying warps as it settles down. In this interpretation we infer
an increase in inclination angle from $32.9\grad$ to $47.9\grad$
between $r=0.25''$ to $1.0''$. The BH mass in the models of
Section~\ref{ss:dyn_mod} scales as $\Mbh \propto \sin^{2}i$ due to the
projection of the rotational velocities. Hence the inferred
differences in inclination angle would amount to an increase in $\Mbh$
by only a factor $\sim 1.5$, which would not significantly change our
results.

IC 1459 hosts an unresolved blue nuclear point source. Carollo
\etal (1997) estimated a luminosity of $L_V \approx 1.5 \times 10^7
\Lsun$. So far, we have assumed that this light is non-stellar radiation 
from the active nucleus. However, one could assume alternatively that
the blue light is emitted by a cluster of young stars. If the cluster
were to have a mass equal to the mass $\Mbh$ that we have inferred
from our models, this would require $\Upsilon_V \gta 10$. Depending on
age and metallicity, stellar evolution models typically predict
$\Upsilon_V \lta 2$ for a young cluster (e.g., Worthey 1994). Thus the
assumption that the point source is stellar in origin cannot lift the
need for a central concentration of non-luminous matter, most likely a
BH.

The data show differences between the fluxes and kinematics for the
various line species. The main characteristics of the observed
kinematics are similar for all species, but we see that for the
species with higher critical densities the flux distribution is
generally more concentrated towards the nucleus and the observed
velocities are higher. This can be well understood qualitatively in
the framework of a single $\Mbh$. However, to actually verify
quantitatively whether each species implies the same value for $\Mbh$
would require information on the flux distributions for each species
at high spatial resolution as well as detailed knowledge on the
ionization structure of the gas, neither of which is available. An
extremely conservative upper limit on the differences in the $\Mbh$
implied by the different line species is obtained by assuming that all
species have the same flux distribution (not actually correct), in
which case $\Mbh$ values are obtained that are up to $\sim 4$ times
larger than inferred from {\Hbeta} and {\HalphaNII}.

Irregularities in the velocity dispersion profiles of {\OIII}, {\OI}
and {\SII} suggest localized turbulent motions. We incorporated
turbulent motion in our models in a very simple manner
(cf.~equation~\ref{eq:turbdef}), using a parametrization that fits the
main trend of an increase of the velocity dispersion toward the
nucleus. Nevertheless, going to the extreme, one could assume that all
observed motions have a non-gravitational origin. The overall
kinematics would then be due to in- or outflows. There are several
objections to this interpretation. A spherical in- or outflow could
not produce any net mean velocity. A bi-directional flow is unlikely
since no hint of this is seen in the HST {\HalphaNII} emission image.

Next we consider the location of IC 1459 with respect to the
correlations between $\Mbh$ and host total optical and radio
luminosity, mentioned in the Introduction. The ratio of $\Mbh$ and
galaxy mass is in the range $0.4$--$1.5 \times 10^{-3}$. This is
somewhat lower than the average value of $\sim 2 \times 10^{-3}$ seen
for other galaxies (Kormendy \& Richstone 1995), but still comfortably
within the observed scatter. As discussed in the Introduction, IC 1459
is probably the end-product of a merger between two galaxies.
Apparently, this merger history has not moved IC 1459 to an atypical
spot in the $\Mbh$ vs.~galaxy mass scatter diagram. The fact that IC
1459 has a LINER type spectrum and core radio emission, but no jets,
makes it interesting to see where it is located on a $\Mbh-L_{\rm
radio}$ plot. IC 1459 has a radio luminosity of $5.5
\times 10^{22} {\rm WHz^{-1}}$ at 5 GHz (Wright \etal 1996).  Interestingly,
this puts IC 1459 within the scatter observed around the correlation
between $\Mbh$ and total radio luminosity inferred for a small sample
of galaxies with available BH mass determinations, but quite off the
correlation with core radio luminosity (see Figs.~3 and 4 of
Franceschini, Vercellone, \& Fabian 1998). However, one should be
weary of beaming and variability of core radio sources, and possible
resolution differences among the observations.

Our dynamical modeling has used the gas disk in IC 1459 as a
diagnostic tool to constrain $\Mbh$. A logical next step to improve on
our work would be to obtain better two-dimensional coverage of the
gaseous and stellar kinematics. A definite advance in our
understanding of nuclear gas disks would also be obtained if we had a
better knowledge of properties of the gas disk such as the electron
density, metallicity, and the ionization structure and ionization
mechanism. One could then try to simultaneously understand these
properties and the gas dynamics. We now know that quite likely most,
or even all, bright ellipticals host a massive central BH. More
detailed knowledge of the chemical properties and kinematics of the
dust and gas surrounding the BH could tell us about the origin of this
material: accretion of small satellites, stripping of companions, or
internal mass loss from stars. This would immediately constrain the
frequency and probability with which any bright elliptical hosts this
material. A better understanding of the kinematics, such as the
importance of dissipative shocks generated by turbulence, could then
help to determine the accretion rate of the black hole. These pieces
of information together could tell us what fraction of the observed
$\Mbh$ could have come from this process over the life time of a
galaxy. Knowledge on the ratio of black hole and stellar mass as
function of time would be valuable for understanding the formation and
evolution of early-type galaxies in general.


\acknowledgments

Support for this work was provided by NASA through grant number
\#GO-06537.01-95A, and through C.M.C.'s Hubble Fellowship
\#HF-01079.01-96A, awarded by the Space Telescope Science Institute which 
is operated by the Association of Universities for Research in
Astronomy, Incorporated, under NASA contract NAS5-26555. The authors
would like to thank Marijn Franx for helpful comments on an earlier
version of the manuscript.


\clearpage



\ifsubmode\else
\baselineskip=10pt
\fi


\clearpage


\ifsubmode\else
\baselineskip=14pt
\fi


\newcommand{\figcapimages}
{{\bf (Left)}: Stellar continuum image of the central $5'' \times 5''$
of IC 1459. The image was obtained through a narrow-band filter,
F631N, which at the redshift of IC 1459 measures primarily stellar
continuum.  The image was rotated such that the galaxy major axis runs
horizontal. A small dust lane runs across the central region. The
location of the FOS apertures inside the $1'' \times 1''$ box drawn at
the center of the image is shown in Fig.~\ref{f:aperpos}. {\bf
(Right)}: {\HalphaNII} emission image, at the same scale and
orientation. There is a disk of gas with an irregular morphology at $r
\gta 0.5''$, with filaments extending in various
directions.\label{f:images}}

\newcommand{\figcapsbprof}
{The stellar surface brightness profile of IC 1459 as a function of
radius along the major axis. The profile, shown with open dots, was
determined by Carollo \etal (1997) from isophotal fits to their
HST/WFPC2 $I$-band image. The solid line shows the best fit model for
the surface brightness using the parameterization of the luminosity
density given in eq.~(\ref{e:lumdendef}). The vertical lines at
$0.15''$ and $5''$ mark the region included in the fit. The exclusion
of the central point source and observations outside $5''$ are
discussed in Section \ref{ss:mass_stars}.\label{f:sbprof}}

\newcommand{\figcapaperpos}
{Aperture positions for the FOS spectra. The $(x,y)$ coordinate system
is centered on the galaxy, with the $x$-axis in the direction of ${\rm
PA} = 30^{\circ}$ (approximately the photometric major axis). The $1''
\times 1''$ box around the diagram corresponds to the central box
drawn in Fig.~\ref{f:images}a.  The ID of each aperture is as in
Table~\ref{t:FOSsetup}, and is used in the remainder of the
paper.\label{f:aperpos}}

\newcommand{\figcapemlines}
{Emission lines in the FOS spectra of IC 1459, with identifications
given in the fourth row of panels. The abscissa is the observed vacuum
wavelength in {\AA}, and the ordinate is the flux in
$\ergscm\>${\AA}$^{-1}$ observed through the aperture (note the
different scale in each panel). Each row of panels corresponds to the
aperture position listed in the left-most panel, as defined in
Fig.~\ref{f:aperpos} and Table~\ref{t:FOSsetup}. Thin dotted curves
are the observed, continuum-subtracted spectra, smoothed with a
Gaussian with a dispersion of $1${\AA} to reduce the noise. The
instrumental resolution is $4.15${\AA} FWHM. The corresponding
dispersion (FWHM/$2.355$) in velocity ranges from $108 \kms$ for
{\Hbeta} to $77 \kms$ for {\SII}. Heavy curves are fits obtained under
the assumption that all emission lines are single Gaussians, as
described in the text. The kinematical quantities $V$ and $\sigma$ of
the Gaussian fits are shown in Figure~\ref{f:gaskin} and listed in
Table~\ref{t:FOSgaskin}.\label{f:emlines}}

\newcommand{\figcapgaskin}
{Mean velocity and velocity dispersion of the ionized gas in IC 1459,
inferred from Gaussian fits to the emission lines in the FOS spectra
(as listed in Table~\ref{t:FOSgaskin} and shown in
Figure~\ref{f:emlines}). The abscissa in each panel is the major axis
distance of the aperture center. Each column of panels corresponds to
a different line species, as labeled in the top panel. The rows of
panels are from top to bottom: the mean velocities for the
small-aperture observations; the mean velocities for the
large-aperture observations; the velocity dispersions for the
small-aperture observations; and the velocity dispersions for the
large-aperture observations. The horizontal bar in the left panel of
each row indicates the size of the square aperture for that
row.\label{f:gaskin}}

\newcommand{\figcapfluxfit}
{Observed emission-line fluxes (solid dots) for {\HalphaNII} (top row)
and {\Hbeta} (bottom row) as function of distance from the galaxy
center. The fluxes are those inferred from the HST/FOS spectra,
HST/WFPC2 narrow-band image and CTIO spectra, respectively, as
indicated in the top left of each panel. Solid lines show the
predictions of the triple exponential flux distribution
(eq.~[\ref{e:fluxparam}]) that best fits all the data (our standard
model), described in Section~\ref{ss:flux_gas}. Note that the radius
and flux scales (in units of $10^{-14} \ergscm$) along the ordinate
and abscissa are different for each panel.\label{f:fluxfit}}

\newcommand{\figcapfluxprof}
{Models of the form given by equation~(\ref{e:fluxparam}) for the
intrinsic flux density distribution as function of radius for
{\HalphaNII} and {\Hbeta}. The solid curve is our standard flux
density model, which was obtained by simultaneously fitting all fluxes
from the WFPC2, FOS and CTIO data (see Figure~\ref{f:fluxfit}). The
dotted curve is the fit to the FOS and CTIO subset of the data, and the
dashed curve is the fit to the WFPC2 and CTIO subset. The dynamically
inferred BH mass does not depend sensitively on which of these flux
distributions is used in the modeling (cf.~Section~\ref{ss:bestfit}).
The fluxes given along the ordinate are for {\HalphaNII}. Fluxes for
{\Hbeta} can be computed using the ratio
F({\Hbeta})/F({\HalphaNII})=0.043 (cf.~Section~\ref{ss:flux_gas}).
\label{f:fluxprof}}

\newcommand{\figcapmodelfos}
{Data-model comparison for the mean velocity (left column) and the
velocity dispersion (right column) of the ionized gas, as inferred from
Gaussian fits to the {\HalphaNII} (open circles) and {\Hbeta} (filled
dots) emission lines. The connected asterisks are the predictions of
the standard kinematical model described in Section~\ref{ss:bestfit},
which has a central BH of mass $\Mbh = 1.0 \times 10^8 \Msun$.  The
top panels are for the small FOS apertures, and those on the
bottom for the large apertures.\label{f:modelfos}}

\newcommand{\figcaprotfits}
{Kinematical predictions of models with various BH masses for the
{\HalphaNII} (open circles) and {\Hbeta} (filled dots) HST rotation
velocity measurements. The heavy solid line shows the predictions of
the standard kinematical model described in Section~\ref{ss:bestfit},
which has a central BH of mass $\Mbh = 1.0 \times 10^8 \Msun$. Dotted
and dashed curves show the predictions of models with $\Mbh = 0$ and
$\Mbh = 7.0 \times 10^8 \Msun$, respectively (with the other input
parameters varied to optimize the fit). These models predict a
velocity gradient that is either too shallow or too steep. Both models
are ruled out at more than 99\% confidence
(cf.~Figure~\ref{f:mbhrange}).\label{f:rotfits}}

\newcommand{\figcapmbhrange}
{Models were constructed with the BH mass fixed to the value shown
along the abscissa. The remaining parameters in each model were chosen
to optimize the fit to the available gas kinematical data for
{\HalphaNII} and {\Hbeta}. The curves in the left panel show for the
two line species the quantity $\chi^2_V$ that measures the quality of
the fit to the HST rotation velocity measurements ({\HalphaNII}: solid
curve; {\Hbeta}: dashed curve). The minimum of $\chi^2_V$ indicates
the best-fitting BH mass. The inferred BH mass is similar for both
emission lines. The curves in the right panel show the corresponding
$\Delta\chi^2_V \equiv \chi^2_V - \chi^2_{V,{\rm min}}$. The dotted
horizontal lines show the $\Delta\chi^2_V$ values that yield the
indicated confidence intervals on the best-fitting BH
mass.\label{f:mbhrange}}

\newcommand{\figcapmodelctio}
{Data points show the gas kinematics of {\Hbeta} inferred from the
ground-based CTIO data. The left panel shows the mean velocity and the
right panel the velocity dispersion. The connected asterisks are the
predictions of the standard kinematical model described in
Section~\ref{ss:bestfit}, which has a central BH of mass $\Mbh = 1.0
\times 10^8 \Msun$. This model was obtained by fitting to the HST data
only. The discrepancies between the predictions of this model and the
CTIO data suggest that the gas may have a certain amount of asymmetric
drift, as discussed in
Section~\ref{ss:CTIOcomp}.\label{f:modelctio}}

\newcommand{\figcapspeciesflux} 
{The relative surface brightnesses ${\cal F} {\rm (aperture,species)}$
(defined in eq.~[\ref{eq:relfluxdef}]) for the different FOS apertures
and line species. Left panel: {\tt 0.1-PAIR} aperture observations;
Right panel: {\tt 0.25-PAIR} aperture observations. Results are shown
for {\OIII} (asterisks), {\OI} (diamonds), {\HalphaNII} (solid dots),
{\Hbeta} (triangles) and {\SII} (crosses). For the small aperture, the
central peakedness of the brightness profile differs strongly for the
different species. The profiles for {\OI} and {\OIII} are more
centrally peaked than that for {\HalphaNII}, while the profile for
{\SII} is less centrally peaked than that for {\HalphaNII}.
\label{f:speciesflux}}

\newcommand{\figcapcritical}
{{\bf (a)} (Left): The velocity dispersion $\sigma {\rm (S1)}$
measured through aperture S1 as function of critical density, for the
different line species. {\bf (b)} (Right): $\sigma {\rm (S1)}$ versus
the flux-peakedness parameter ${\cal P}$ (defined in
eq.~[\ref{eq:peakedness}]). Line species are plotted with the same
symbols as in Fig.~\ref{f:speciesflux} (but in panel (a) the solid dot
refers to {\NII} only). Lines for which the velocity dispersion is
high generally have a high critical density, and have a flux that is
more strongly concentrated towards the center. This has a natural
interpretation in terms of ionization around a central BH, as
discussed in the text.\label{f:critical}}

\newcommand{\figcapstarkin}
{RMS stellar line-of-sight velocities along the major axis of IC 1459,
determined from ground-based observations. The curves are the
predictions of $f(E,L_z)$ models with central BH masses of, from
bottom to top respectively, $\Mbh = 0$, $1.1$, $2.3$, $3.4$, $4.5
\times 10^9 \Msun$. None of the models fits particularly well, 
but the models that fit best have $\Mbh \approx 3$--$4 \times 10^9
\Msun$. This differs from the value inferred from the HST gas kinematics,
presumably, as discussed in the text, because the assumption of a
two-integral distribution function is not appropriate for IC
1459.\label{f:starkin}}


\ifsubmode
\figcaption{\figcapimages}
\figcaption{\figcapsbprof}
\figcaption{\figcapaperpos}
\figcaption{\figcapemlines}
\figcaption{\figcapgaskin}
\figcaption{\figcapfluxfit}
\figcaption{\figcapfluxprof}
\figcaption{\figcapmodelfos}
\figcaption{\figcaprotfits}
\figcaption{\figcapmbhrange}
\figcaption{\figcapmodelctio}
\figcaption{\figcapspeciesflux}
\figcaption{\figcapcritical}
\figcaption{\figcapstarkin}
\clearpage
\else\printfigtrue\fi

\ifprintfig


\clearpage
\begin{figure}
\plottwo{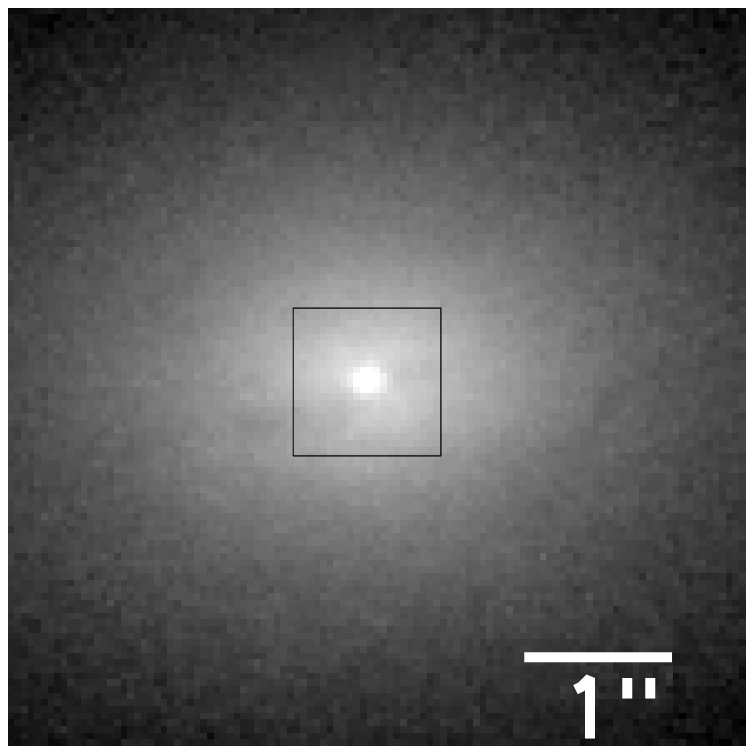}{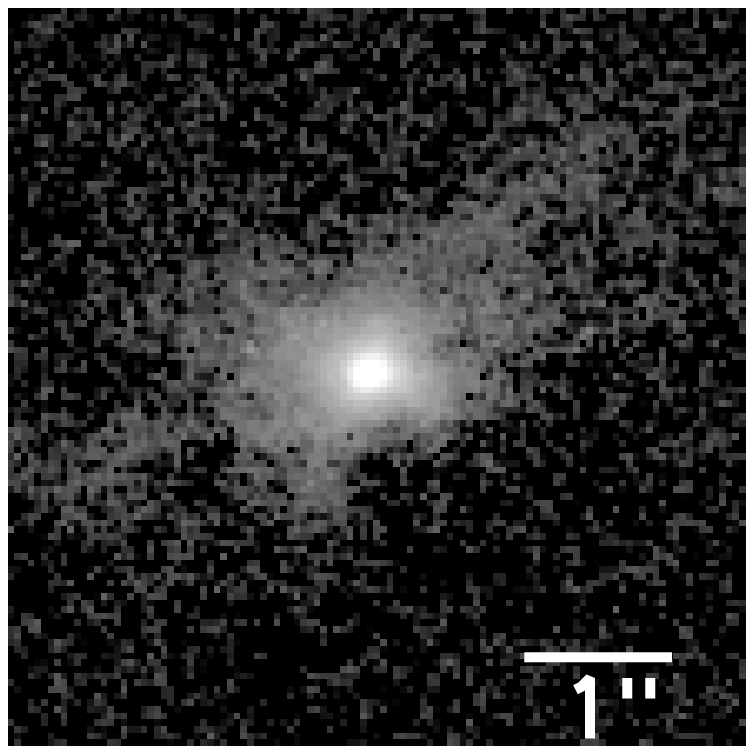}
\ifsubmode
\vskip3.0truecm
\setcounter{figure}{0}
\addtocounter{figure}{1}
\centerline{Figure~\thefigure}
\else\figcaption{\figcapimages}\fi
\end{figure}

 
\clearpage
\begin{figure}
\centerline{\epsfbox{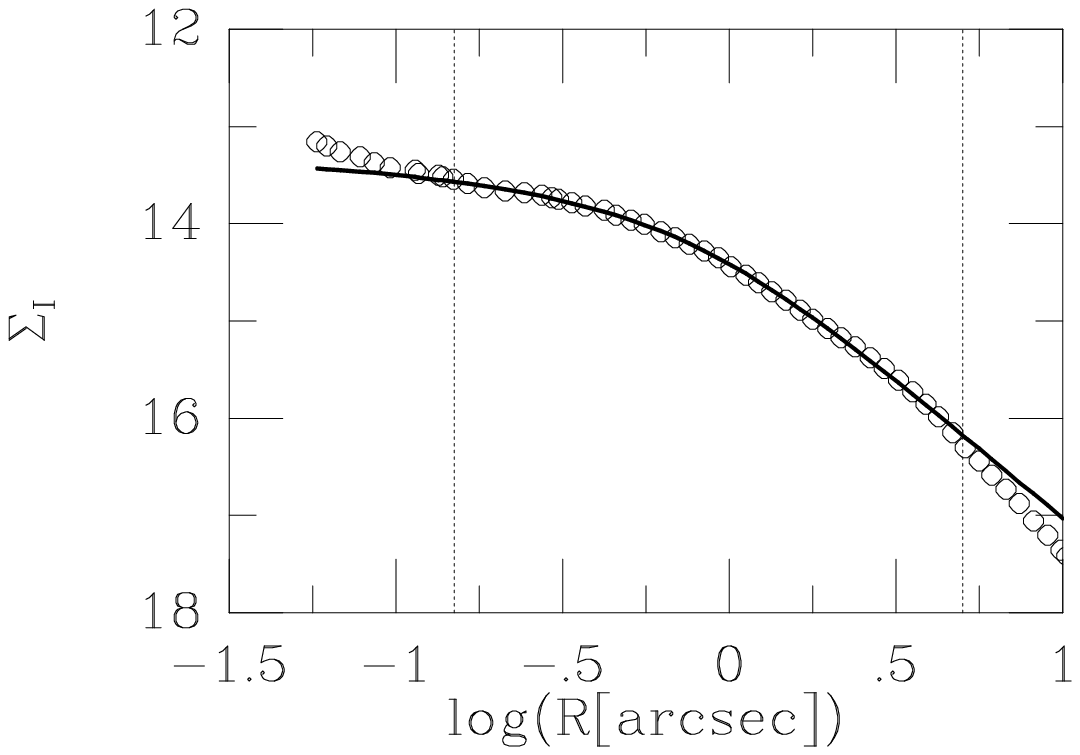}}
\ifsubmode
\vskip3.0truecm
\addtocounter{figure}{1}
\centerline{Figure~\thefigure}
\else\figcaption{\figcapsbprof}\fi
\end{figure}

 
\clearpage
\begin{figure}
\centerline{\epsfbox{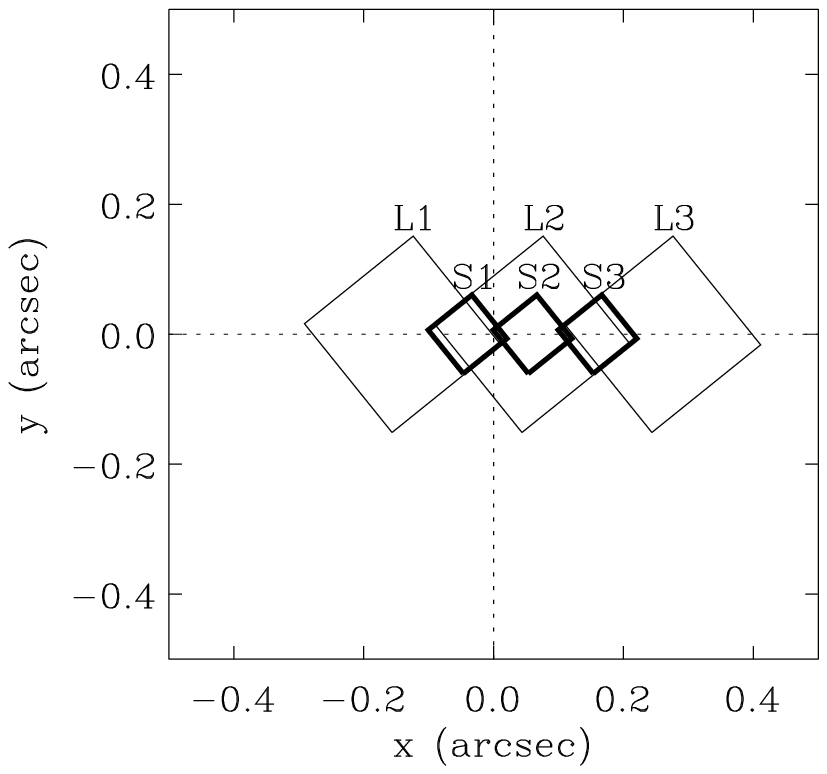}}
\ifsubmode
\vskip3.0truecm
\addtocounter{figure}{1}
\centerline{Figure~\thefigure}
\else\figcaption{\figcapaperpos}\fi
\end{figure}

 
\clearpage
\begin{figure}
\epsfxsize=12.0truecm
\centerline{\epsfbox{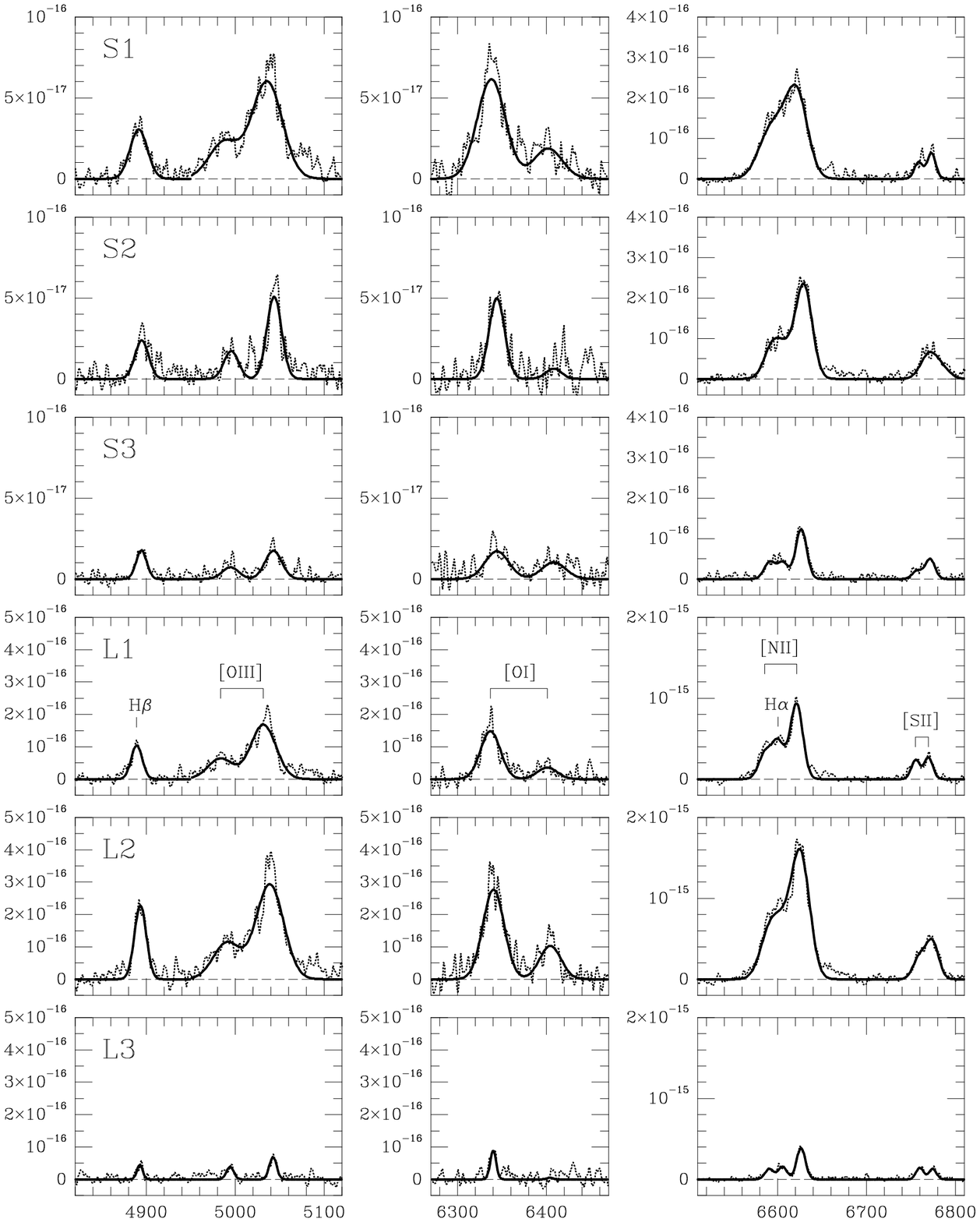}}
\ifsubmode
\vskip3.0truecm
\addtocounter{figure}{1}
\centerline{Figure~\thefigure}
\else\figcaption{\figcapemlines}\fi
\end{figure}
 
 
\clearpage
\begin{figure}
\epsfxsize=12.0truecm
\centerline{\epsfbox{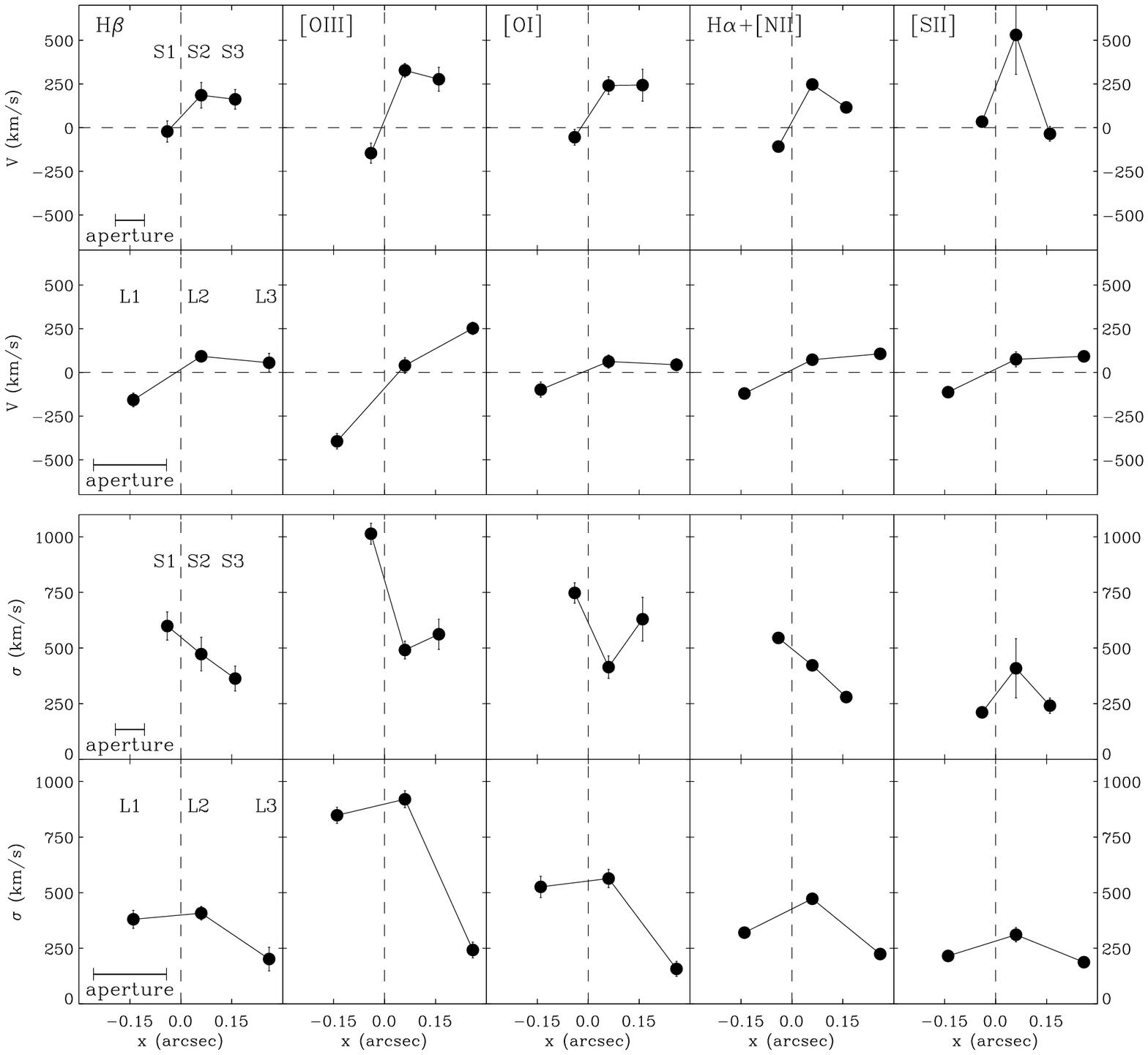}}
\ifsubmode
\vskip3.0truecm
\addtocounter{figure}{1}
\centerline{Figure~\thefigure}
\else\figcaption{\figcapgaskin}\fi
\end{figure}

 
\clearpage
\begin{figure}
\epsfxsize=0.9\hsize
\centerline{\epsfbox{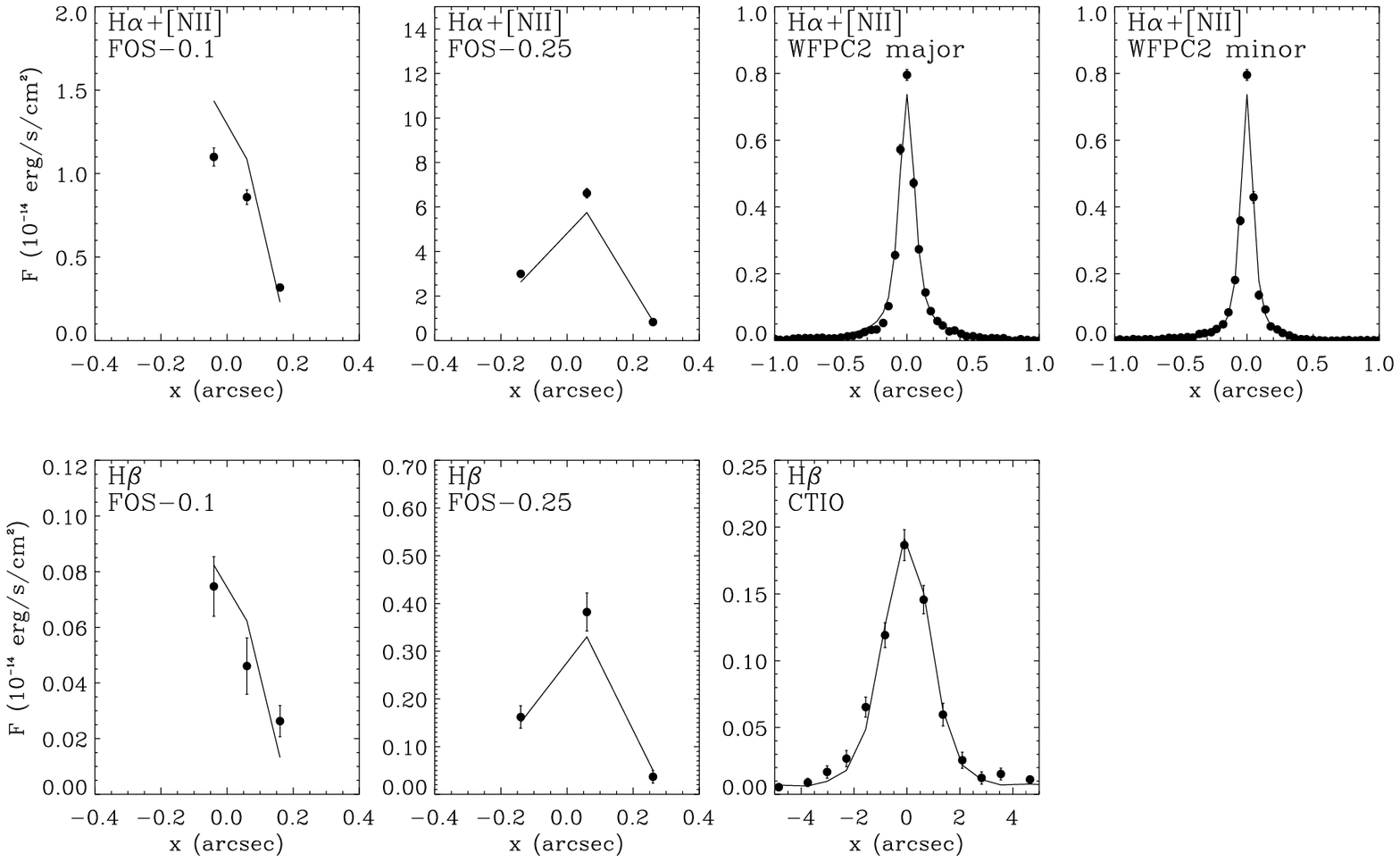}}
\ifsubmode
\vskip3.0truecm
\addtocounter{figure}{1}
\centerline{Figure~\thefigure}
\else\figcaption{\figcapfluxfit}\fi
\end{figure}

 
\clearpage
\begin{figure}
\epsfxsize=9.0truecm
\centerline{\epsfbox{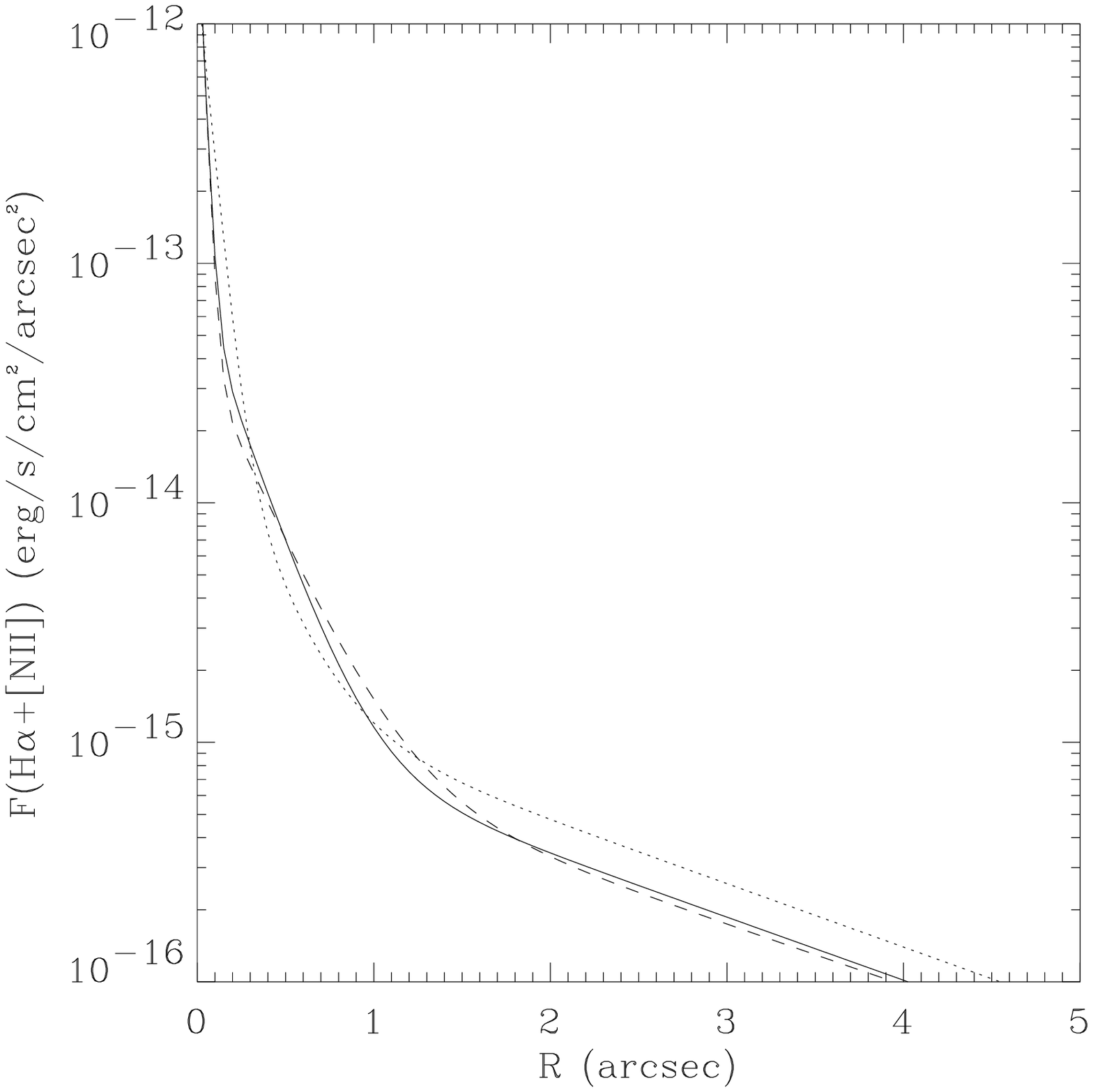}}
\ifsubmode
\vskip3.0truecm
\addtocounter{figure}{1}
\centerline{Figure~\thefigure}
\else\figcaption{\figcapfluxprof}\fi
\end{figure}
 
 
\clearpage
\begin{figure}
\epsfxsize=0.9\hsize
\centerline{\epsfbox{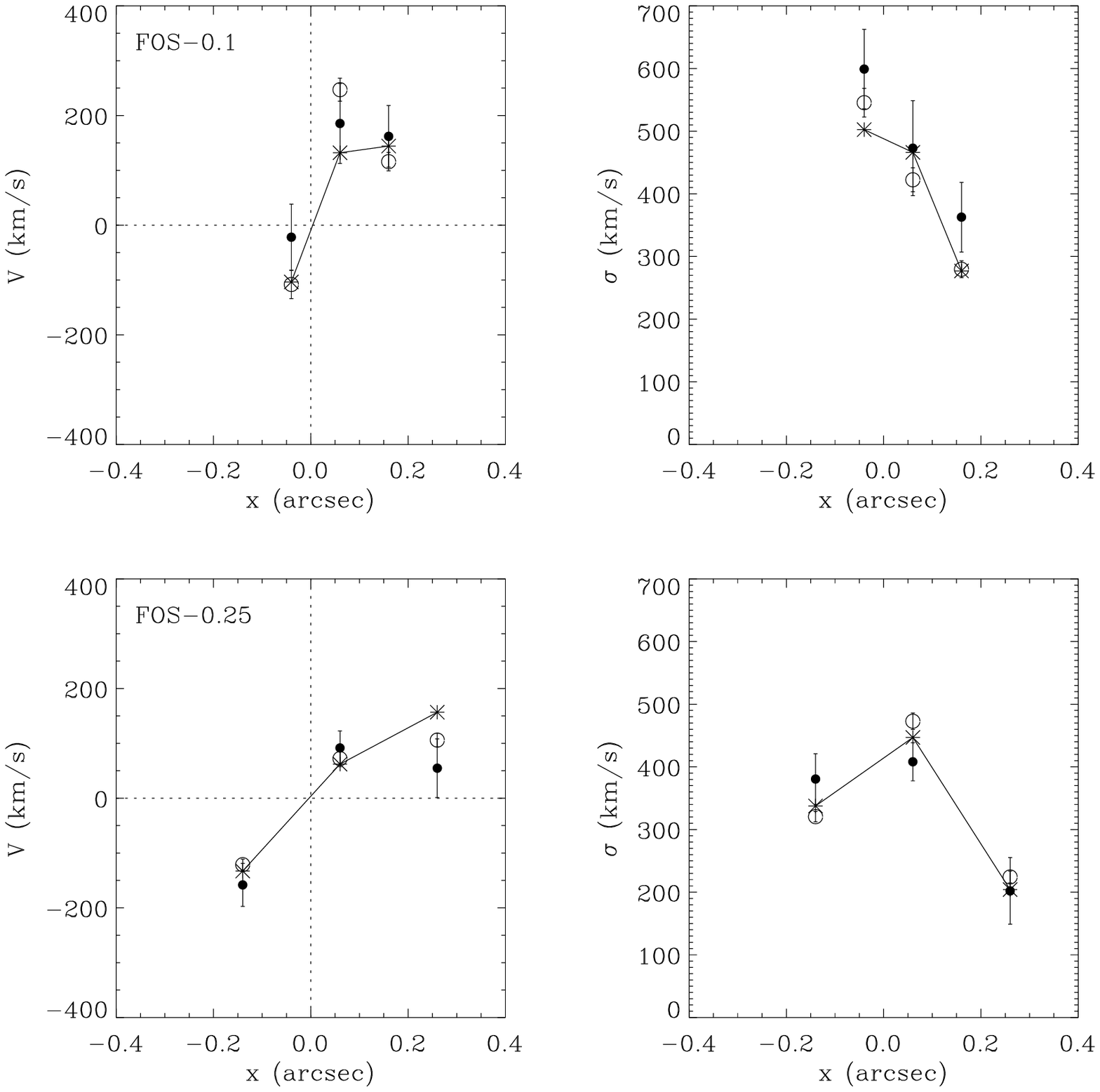}}
\ifsubmode
\vskip3.0truecm
\addtocounter{figure}{1}
\centerline{Figure~\thefigure}
\else\figcaption{\figcapmodelfos}\fi
\end{figure}

 
\clearpage
\begin{figure}
\epsfxsize=0.9\hsize
\centerline{\epsfbox{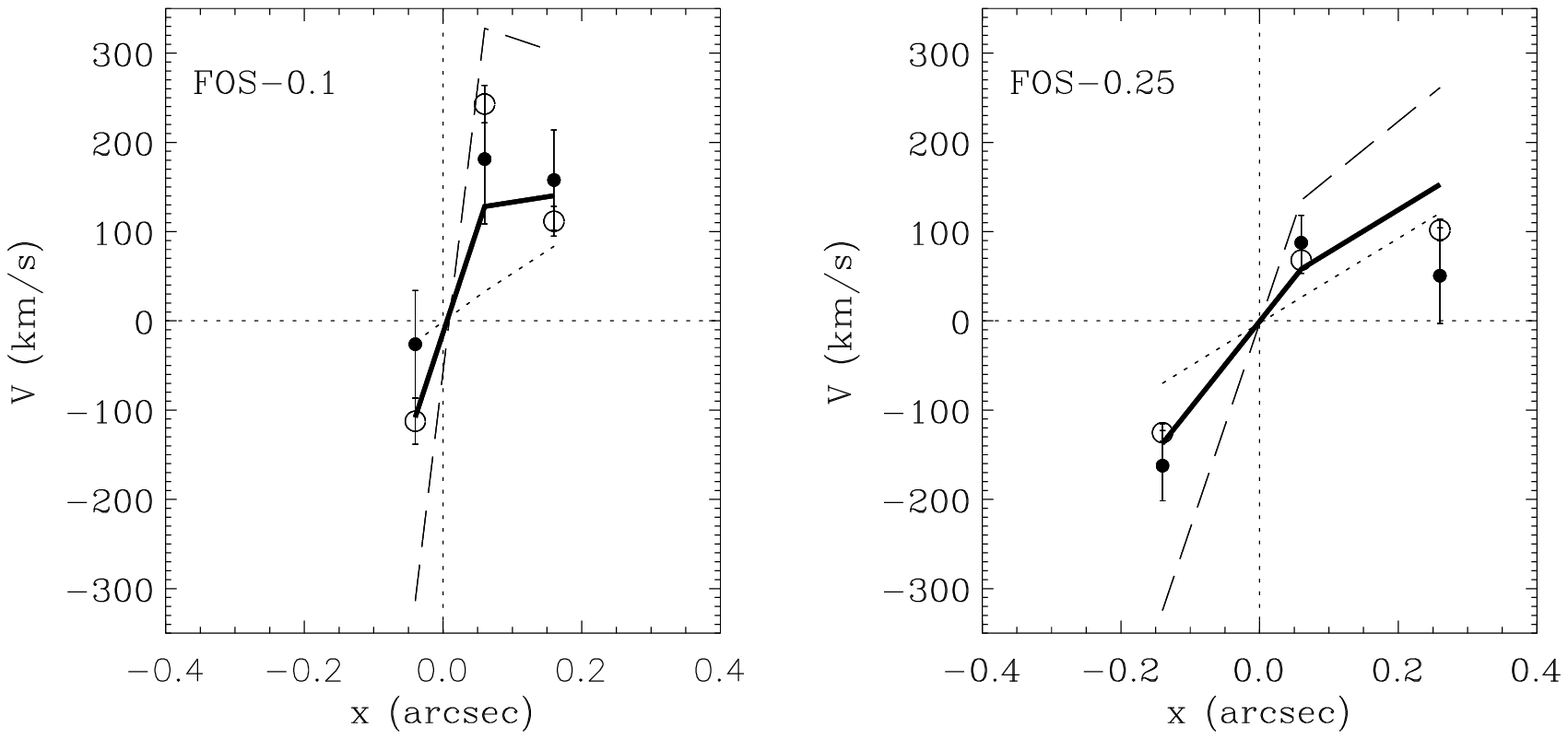}}
\ifsubmode
\vskip3.0truecm
\addtocounter{figure}{1}
\centerline{Figure~\thefigure}
\else\figcaption{\figcaprotfits}\fi
\end{figure}

 
\clearpage
\begin{figure}
\epsfxsize=0.9\hsize
\centerline{\epsfbox{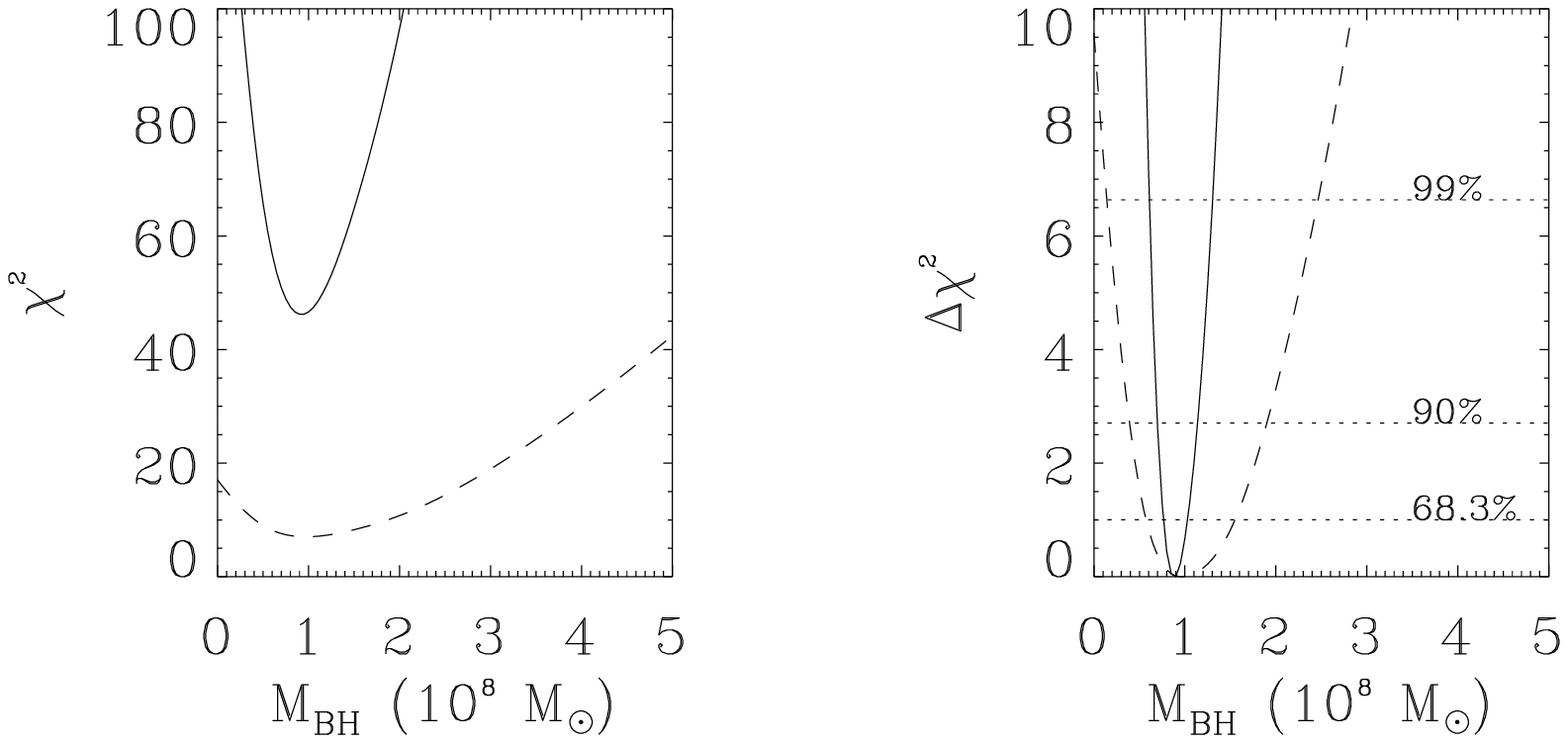}}
\ifsubmode
\vskip3.0truecm
\addtocounter{figure}{1}
\centerline{Figure~\thefigure}
\else\figcaption{\figcapmbhrange}\fi
\end{figure}


\clearpage
\begin{figure}
\epsfxsize=0.9\hsize
\centerline{\epsfbox{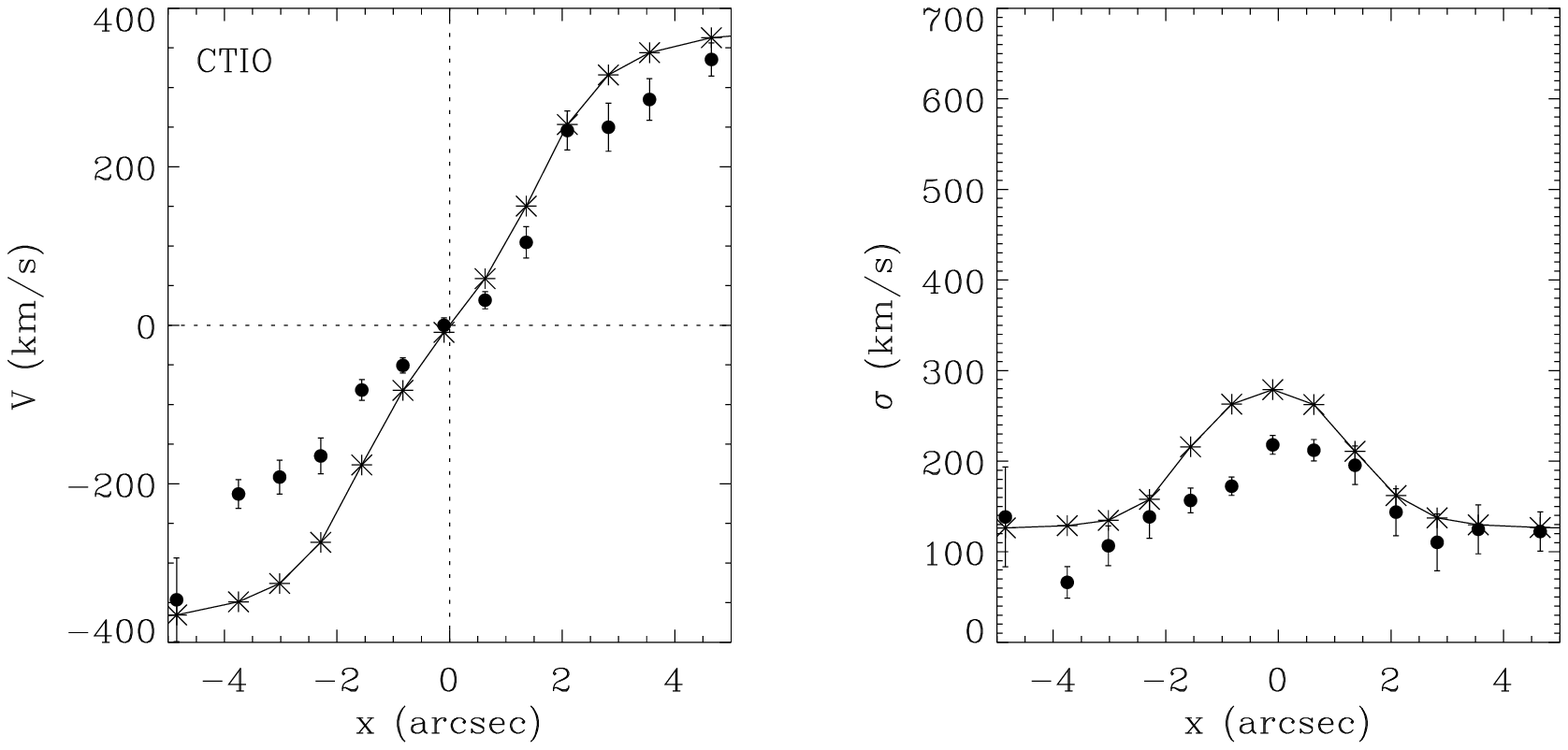}}
\ifsubmode
\vskip3.0truecm
\addtocounter{figure}{1}
\centerline{Figure~\thefigure}
\else\figcaption{\figcapmodelctio}\fi
\end{figure}
 
 
\clearpage
\begin{figure}
\epsfxsize=0.9\hsize
\centerline{\epsfbox{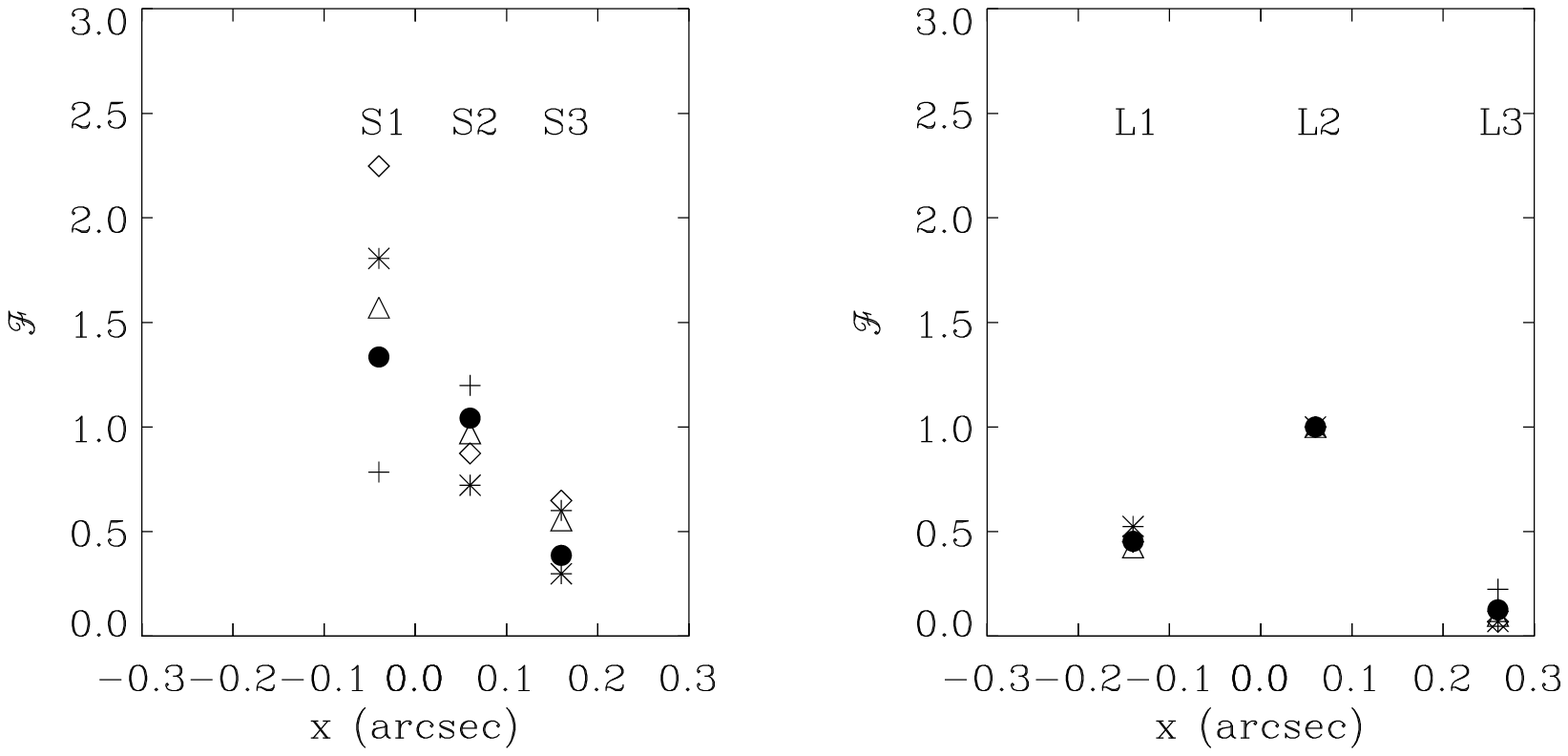}}
\ifsubmode
\vskip3.0truecm
\addtocounter{figure}{1}
\centerline{Figure~\thefigure}
\else\figcaption{\figcapspeciesflux}\fi
\end{figure}
 
 
\clearpage
\begin{figure}
\epsfxsize=0.9\hsize
\centerline{\epsfbox{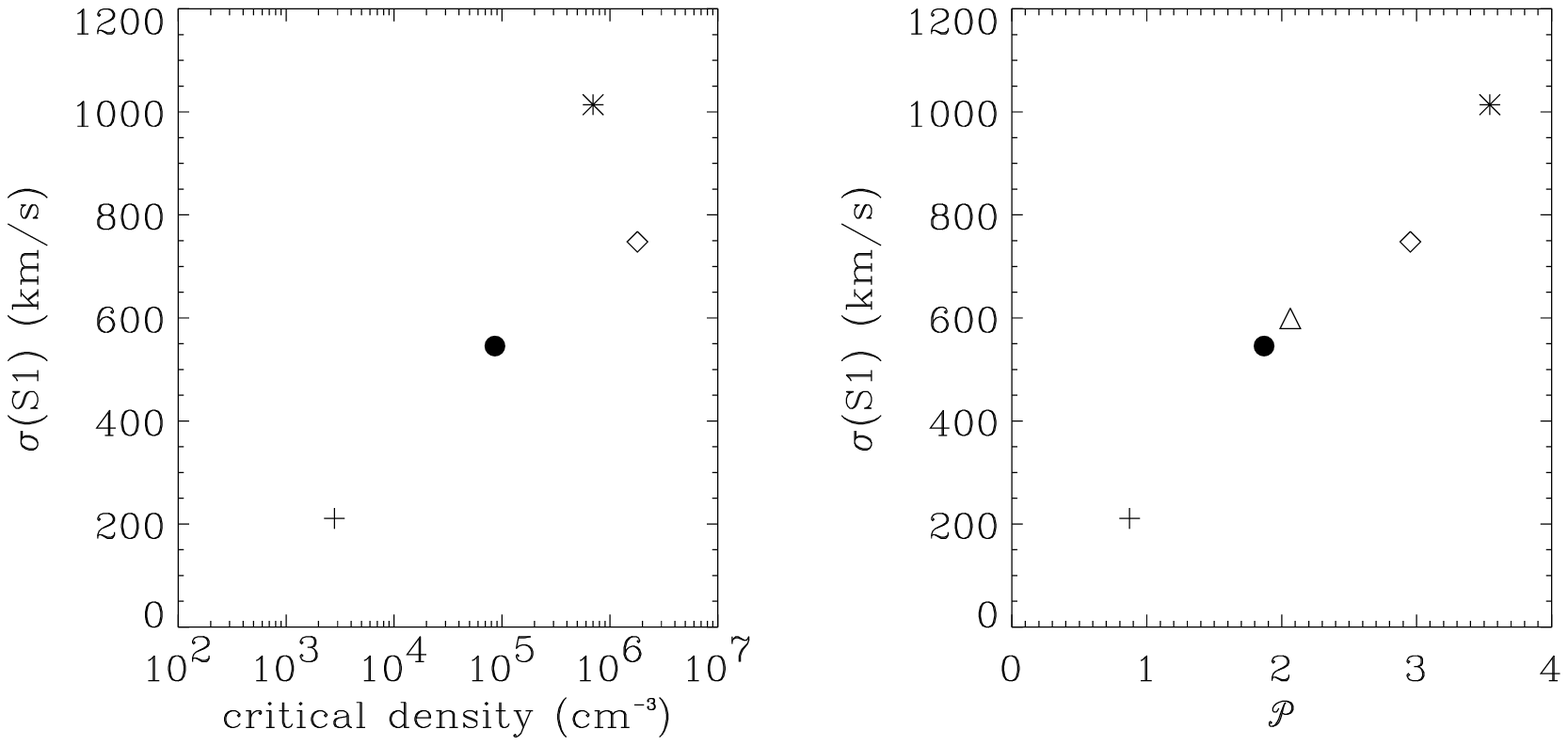}}
\ifsubmode
\vskip3.0truecm
\addtocounter{figure}{1}
\centerline{Figure~\thefigure}
\else\figcaption{\figcapcritical}\fi
\end{figure}

 
\clearpage
\begin{figure}
\centerline{\epsfbox{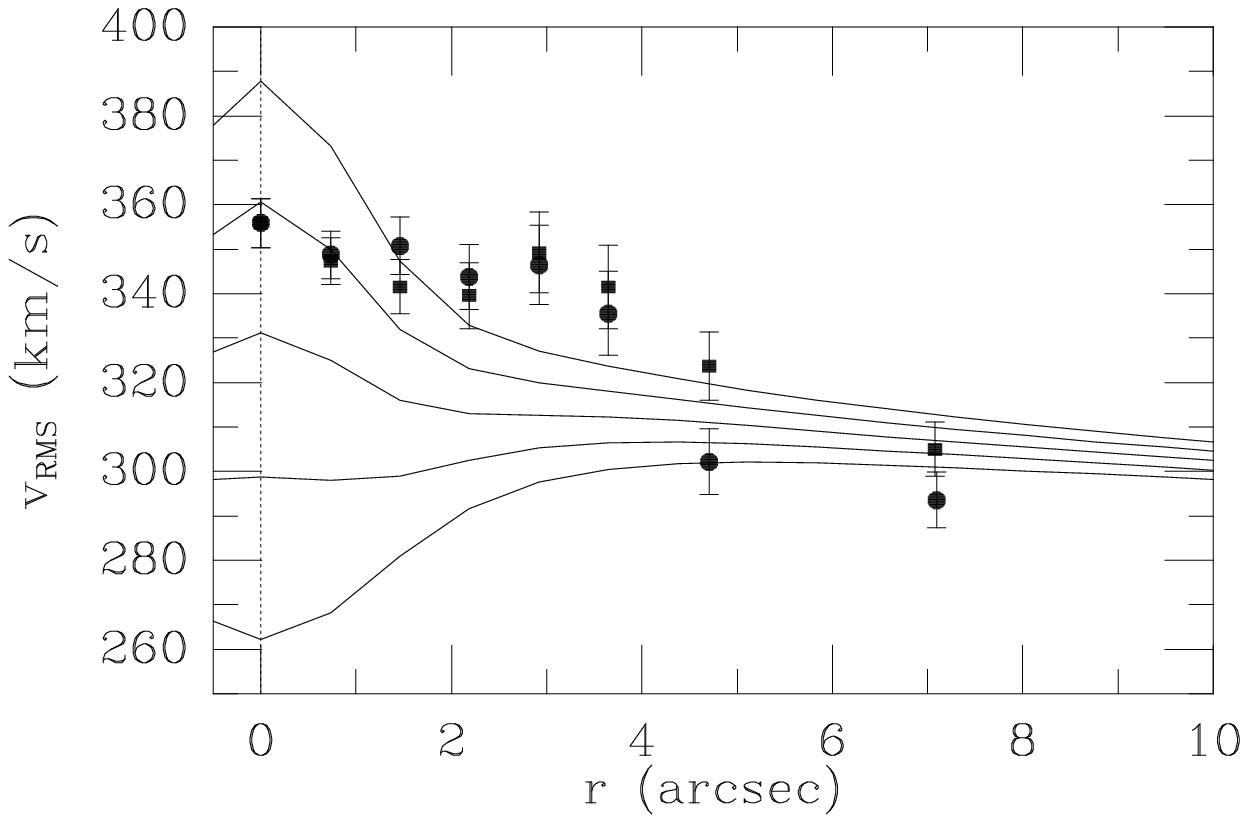}}
\ifsubmode
\vskip3.0truecm
\addtocounter{figure}{1}
\centerline{Figure~\thefigure}
\else\figcaption{\figcapstarkin}\fi
\end{figure}
 

\fi


\clearpage
\ifsubmode\pagestyle{empty}\fi


\begin{deluxetable}{lccc}
\tablewidth{\hsize}
\renewcommand{\tabcolsep}{46pt}
\tablecaption{HST/WFPC2 images: observational setup\label{t:WFPC2}}
\tablehead{
\colhead{filter} & \colhead{$\lambda_0$} & \colhead{$\Delta \lambda$} & 
                   \colhead{$T_{\rm exp}$} \\ & \colhead {(\AA)} &
                 \colhead{(\AA)} & \colhead{(s)} \\
\colhead{(1)} & \colhead{(2)} & \colhead{(3)} & \colhead{(4)} \\
}
\startdata
LRF      & 6611 & 81 & 2000 \\
F631N    & 6306 & 31 & 2300 \\
\enddata
\tablecomments{The filter name is listed in column~(1). Column~(2)
and~(3) list the central wavelength of the filter and the FWHM.
Column~(4) lists the total exposure time, which for each
filter was divided over three different exposures.}
\end{deluxetable}


\clearpage 
\begin{deluxetable}{ccrr}
\tablewidth{\hsize}
\renewcommand{\tabcolsep}{43pt}
\tablecaption{HST/FOS spectra: observational setup\label{t:FOSsetup}}
\tablehead{
\colhead{ID} & \colhead{size} & \colhead{$x$} & 
               \colhead{$T_{\rm exp}$} \\
& \colhead{(arcsec)} & \colhead{(arcsec)} & 
               \colhead{(s)} \\
\colhead{(1)} & \colhead{(2)} & \colhead{(3)} &
                \colhead{(4)} \\
}
\startdata
S1 & 0.086 & $-0.04$ & 2490 \\
S2 & 0.086 & $ 0.06$ & 1620 \\
S3 & 0.086 & $ 0.16$ & 2490 \\
L1 & 0.215 & $-0.14$ & 1140 \\
L2 & 0.215 & $ 0.06$ &  620 \\
L3 & 0.215 & $ 0.26$ & 1130 \\
\enddata
\tablecomments{FOS spectra of IC 1459 were obtained through square
  apertures at six different aperture positions. Column~(1) is the ID
  for the aperture as defined in Fig.~\ref{f:aperpos}. Column~(2) lists
  the nominal size of the aperture; there is some evidence that the
  actual aperture sizes may be smaller by up to $0.02''$ (van der
  Marel \etal 1997). Column~(3) lists the aperture position along the
  major axis for each observation. The zeropoint is at the center of
  the galaxy; positive values lie in the direction of ${\rm PA} =
  30^{\circ}$. The uncertainties in these positions are $\sim
  0.02''$. All apertures were found to be positioned on the major axis
  ($y = 0.00''$) to within the same uncertainties. Column~(4) lists
  the exposure times.}
\end{deluxetable}


\begin{deluxetable}{crrrrr}
\tabletypesize{\footnotesize}
\tablewidth{\hsize}
\renewcommand{\tabcolsep}{20pt}
\tablecaption{HST/FOS spectra: gas kinematics\label{t:FOSgaskin}}
\tablehead{
\colhead{ID} & \colhead{Species} & \colhead{$V$} & 
               \colhead{$\Delta V$} & \colhead{$\sigma$} & 
               \colhead{$\Delta \sigma$} \\ 
& & \colhead{(km~s$^{-1}$)} & 
               \colhead{(km~s$^{-1}$)} & \colhead{(km~s$^{-1}$)} & 
               \colhead{(km~s$^{-1}$)} \\
\colhead{(1)} & \colhead{(2)} & \colhead{(3)} &
                \colhead{(4)} & \colhead{(5)} & \colhead{(6)} \\}
\renewcommand{\arraystretch}{0.9}
\startdata
S1 & {\Hbeta}     &  -22 & 60 &  599 &  63 \\
   & {\OIII}      & -146 & 57 & 1014 &  47 \\
   & {\OI}        &  -55 & 45 &  748 &  46 \\
   & {\HalphaNII} & -108 & 26 &  545 &  23 \\
   & {\SII}       &   34 & 32 &  211 &  25 \\
\\
S2 & {\Hbeta}     &  185 & 73 &  473 &  76 \\
   & {\OIII}      &  328 & 40 &  491 &  40 \\
   & {\OI}        &  241 & 50 &  414 &  50 \\
   & {\HalphaNII} &  247 & 21 &  422 &  19 \\
   & {\SII}       &  531 &226 &  409 & 133 \\
\\
S3 & {\Hbeta}     &  162 & 56 &  363 &  56 \\
   & {\OIII}      &  277 & 68 &  562 &  68 \\
   & {\OI}        &  243 & 91 &  630 &  98 \\
   & {\HalphaNII} &  116 & 17 &  280 &  13 \\
   & {\SII}       &  -35 & 42 &  241 &  35 \\
\\
L1 & {\Hbeta}     &  -158 & 39 &  381 &  40 \\
   & {\OIII}      &  -395 & 44 &  848 &  36 \\
   & {\OI}        &   -99 & 43 &  526 &  48 \\
   & {\HalphaNII} &  -121 & 10 &  321 &   8 \\
   & {\SII}       &  -113 & 22 &  216 &  15 \\
\\
L2 & {\Hbeta}     &    92 & 31 &  408 &  30 \\
   & {\OIII}      &    39 & 45 &  920 &  38 \\
   & {\OI}        &    62 & 38 &  564 &  41 \\
   & {\HalphaNII} &    72 & 15 &  473 &  13 \\
   & {\SII}       &    75 & 43 &  311 &  32 \\
\\
L3 & {\Hbeta}     &    55 & 54 &  202 &  53 \\
   & {\OIII}      &   252 & 35 &  243 &  36 \\
   & {\OI}        &    43 & 35 &  158 &  34 \\
   & {\HalphaNII} &   106 & 17 &  280 &  13 \\
   & {\SII}       &    91 & 26 &  188 &  19 \\
\enddata
\tablecomments{\small FOS spectra of IC 1459 were obtained with two different
apertures, at a total of six different aperture positions. Column~(1)
is the ID for the observation as defined in Fig.~\ref{f:aperpos} and
Table~\ref{t:FOSsetup}. Column~(2) identifies the emission line
species. Columns~(3)-(6) list the mean velocity $V$ and velocity
dispersion $\sigma$ of the emission line gas, with corresponding
formal errors, determined from Gaussian fits to the emission lines as
described in the text.}
\end{deluxetable}


\begin{deluxetable}{crrrrrrrrr}
\tabletypesize{\footnotesize}
\tablewidth{\hsize}
\renewcommand{\tabcolsep}{5pt}
\tablecaption{CTIO spectra: gas kinematics\label{t:CTIO}}
\tablehead{
\colhead{ } & \colhead{ } & \multicolumn{4}{c}{{\Hbeta}} & 
\multicolumn{4}{c}{{\OIII}} \\
\colhead{$r$} & \colhead{rebin} & 
\colhead{$V$} & \colhead{$\Delta V$} & \colhead{$\sigma$} & \colhead{$\Delta 
\sigma$} &
\colhead{$V$} & \colhead{$\Delta V$} & \colhead{$\sigma$} & \colhead{$\Delta 
\sigma$} \\
\colhead{arcsec} & \colhead{ } & 
\colhead{(km~s$^{-1}$)} & \colhead{(km~s$^{-1}$)} & \colhead{(km~s$^{-1}$)} & 
\colhead{(km~s$^{-1}$)} &
\colhead{(km~s$^{-1}$)} & \colhead{(km~s$^{-1}$)} & \colhead{(km~s$^{-1}$)} & 
\colhead{(km~s$^{-1}$)} \\
\colhead{(1)} & \colhead{(2)} & 
\colhead{(3)} & \colhead{(4)} & \colhead{(5)} & \colhead{(6)} &
\colhead{(7)} & \colhead{(8)} & \colhead{(9)} & \colhead{(10)} \\
}
\startdata
  4.65 &  2 & 335 &   21 &  122 &   22  &  343 &  10 & 125 &  10 \\
  3.55 &  1 & 284 &   26 &  125 &   27  &  313 &  15 & 154 &  15 \\
  2.82 &  1 & 249 &   30 &  110 &   31  &  301 &  12 & 132 &  12 \\
  2.09 &  1 & 245 &   25 &  144 &   26  &  265 &  17 & 185 &  17 \\
  1.36 &  1 & 104 &   20 &  195 &   21  &  104 &  15 & 261 &  16 \\
  0.63 &  1 &  31 &   11 &  212 &   12  &   47 &   9 & 296 &   9 \\
 -0.10 &  1 &   0 &   10 &  218 &   10  &    0 &   7 & 320 &   8 \\
 -0.83 &  1 & -50 &   10 &  172 &   10  &  -63 &   9 & 285 &   9 \\
 -1.56 &  1 & -81 &   13 &  157 &   14  & -130 &  10 & 183 &  10 \\
 -2.29 &  1 & 164 &   23 &  138 &   24  & -159 &  13 & 170 &  13 \\
 -3.02 &  1 & 191 &   21 &  107 &   22  & -223 &  15 & 155 &  15 \\
 -3.75 &  1 & 212 &   18 &   66 &   17  & -267 &  19 & 133 &  19 \\
 -4.85 &  2 & 346 &   53 &  138 &   55  & -315 &  21 & 151 &  22 \\
\enddata
\tablecomments{\small Kinematics of {\Hbeta} and {\OIII} inferred from
  a long-slit major axis spectrum of IC 1459 obtained at the CTIO 4m
  telescope. Individual pixels along the $1.5''$-wide slit are
  $0.73''$ in size. Column~(1) gives the distance of the aperture
  center from the galaxy center in arcseconds (measured along the
  major axis). Some spatial rebinning along the slit was performed to
  increase the $S/N$. Column~(2) gives the number of binned pixels.
  Columns~(3)--(6) and Columns~(7)--(10) list the mean velocity $V$ and
  velocity dispersion $\sigma$ of the emission line gas, with
  corresponding formal errors, for {\Hbeta} and {\OIII}, respectively,
  determined from single Gaussian fits to the emission lines.}
\end{deluxetable}



\end{document}